\documentclass[onecolumn,showpacs,preprintnumbers,amsmath,amssymb]{revtex4}

\usepackage{subfigure,dcolumn,jabbrv}
\usepackage{graphicx}
\usepackage{dcolumn}
\usepackage{bm}
\usepackage{CJK}
\usepackage{url}
\usepackage{color}
\usepackage{booktabs}
\usepackage{natbib}
\usepackage{hyperref}
\usepackage{cleveref}

\begin{document}


\title{Impact of The Newly Revised Gravitational Redshift of X-ray Burster GS 1826-24 \\ on The Equation of State of Supradense Neutron-Rich Matter}



\author{Wen-Jie Xie$^{1,4}$$\footnote{wenjiexie@yeah.net}$, Bao-An Li$^{2}$$\footnote{Corresponding Author: Bao-An.Li@Tamuc.edu}$, and Nai-Bo Zhang$^{3}$$\footnote{naibozhang@seu.edu.cn}$}
\address {$^{1}$Department of Physics, Yuncheng University, Yuncheng 044000, China}
\address{$^2$Department of Physics and Astronomy, Texas A$\&$M University-Commerce, Commerce, TX 75429-3011, USA}
\address{$^3$School of Physics, Southeast University, Nanjing 211189, China}
\address{$^4$Guangxi Key Laboratory of Nuclear Physics and Nuclear Technology, Guangxi Normal University, Guilin 541004, China}

\date{\today}

\begin{abstract}
Thanks to the recent advancement in producing rare isotopes and measuring their masses with unprecedented precision, the updated nuclear masses around the waiting-point nucleus $^{64}$Ge in the rapid-proton capture process have led to a significant revision of the surface gravitational redshift of the neutron star (NS) in GS 1826-24 by re-fitting its X-ray burst light curve ({\it X. Zhou et al., Nature Physics {\bf 19}, 1091 (2023)}) using Modules for Experiments in Stellar Astrophysics (MESA).
The resulting NS compactness $\xi$ is between 0.183 and 0.259 at 95\% confidence level and its upper boundary is significantly smaller than the maximum $\xi$ previously known. Incorporating this new data within a comprehensive Bayesian statistical framework, we investigate its impact on the Equation of State (EOS) of supradense neutron-rich matter and the required spin frequency for GW190814's minor $m_2$ with mass $2.59\pm 0.05$M$_{\odot}$ to be a rotationally stable pulsar. We found that the EOS of high-density symmetric nuclear matter (SNM) has to be softened significantly while the symmetry energy at supersaturation densities stiffened compared to our prior knowledge from earlier analyses using data from both astrophysical observations and terrestrial nuclear experiments. In particular, the skewness $J_0$ characterizing the stiffness of high-density SNM decreases significantly, while the slope $L$, curvature $K_{\rm{sym}}$, and skewness $J_{\rm{sym}}$ of nuclear symmetry energy all increase appreciably compared to their fiducial values. We also found that the most probable spin rate for the $m_2$ to be a stable pulsar is very close to its mass-shedding limit once the revised redshift data from GS 1826-24 is considered, making the $m_2$ unlikely the most massive NS observed so far.
\end{abstract}

\maketitle

\section{Introduction}
A neutron star (NS) of mass $M$ and radius $R$ has a compactness $\xi\equiv M/R$ (adopting $c=G=1$) and a surface gravitational redshift $z$ defined by $1+z\equiv (1-2\xi)^{-1/2}$. It has long been expected that observational constraints on the redshift $z$ and thus the compactness $\xi$ will provide invaluable information about the Equation of State (EOS) of supradense neutron-rich matter in NSs. However, little progress has been made so far in this direction mostly due to various uncertainties involved in both observations (e.g. distance $d$ to the source of X-ray bursters) and some ingredients (e.g. masses of nuclei involved in rapid-proton (rp) capture processes) of model simulations of the relevant astrophysical processes, for reviews, see e.g., Refs. \cite{Schatz:2022vzq,Lattimer:2006xb}. Nevertheless, significant efforts have been devoted to improving the situation in recent years by reducing or removing the uncertainties in both astrophysical observations and nuclear physics inputs for their simulations. For example,
GS 1826-24 is a unique Type-I X-ray burster powered by the rp process involving hundreds of exotic neutron-deficient nuclides. While remarkable agreements between observations and theoretical models of recurrence times, energetics, and light curves have been found in several comprehensive analyses, see, e.g., Refs. \cite{Galloway2004,Heger2007,Zand2009,Zamfir2011,Meisel2019}, our poor knowledge about the masses of nuclides around the so-called waiting points in the rp process has been among the major uncertainties in simulating the light curves of GS 1826-24. The masses of these nuclides play a decisive role in matter flow and therefore the produced X-ray flux. Because these nuclides are very short-lived, it is very tough to measure their masses precisely. In particular, $^{64}$Ge is among a handful of the rp process waiting-point nuclides. Fortunately, thanks to the technical advancement and breakthroughs at the Lanzhou/China complex \cite{Xia2002,Zhan2010} of rare isotope facility (for producing short-lived nuclides) and cooler-storage ring (used as a mass spectrometer) \cite{Tu:2011zz,Zhang:2012ss}, precision measurements of the masses of $^{63}$Ge, $^{64,65}$As and $^{66,67}$Se-the relevant nuclear masses around the waiting-point $^{64}$Ge have been achieved for the first time recently \cite{zhouMassMeasurementsShow2023a}. The high-precision masses of these nuclides have been used as inputs in simulating the X-ray burst from GS 1826-24 using Modules for Experiments in Stellar Astrophysics (MESA) \cite{Paxton:2015jva}. By fitting its light curve, its optimal distance $d$ (redshift $z$) was found to increase (decrease) by about 6.5\% (6.3\%) \cite{zhouMassMeasurementsShow2023a} compared to analyses using the 2020 atomic mass table AME'20 \cite{Wang:2021xhn}. In particular, the new analyses incorporating the high-precision masses from Lanzhou yield a compactness $\xi$ ranging from approximately 0.183 to 0.259 at a 95\% confidence level \cite{Cai:2023pkt}. Its upper boundary is significantly smaller than the maximum $\xi$ previously known \cite{Lindblom84}. Compared to the results of analyses using AME'20, the Lanzhou results suggest a less-compact NS in GS 1826–24, indicating that an adjustment of the NS matter EOS may be necessary \cite{zhouMassMeasurementsShow2023a}. However, it is unclear what aspects of the EOS and how much they may be affected by the newly revised redshift
of the NS in GS 1826–24 compared to our current knowledge about the nuclear EOS.

Generally speaking, a reduction of the NS compactness $\xi\equiv M/R$ requires either reducing its mass $M$ (which then requires particularly the softening of symmetric nuclear matter (SNM) EOS at supersaturation densities) and/or enlarging its radius $R$ (which then requires the stiffening of nuclear symmetry energy especially at supersaturation densities). In this work, within a comprehensive Bayesian framework besides using existing astrophysical constraints on the masses and radii of several NSs from LIGO/VIRGO, NICER, XMM-Newton, and Chandra we incorporate the newly revised redshift $z$ for NS in GS 1826-24 and investigate its impact on:
\begin{enumerate}
\item Parameters characterizing the EOS of symmetric nuclear matter (e.g. its incompressibility $K_0$ and skewness $J_0$ around its saturation density $\rho_0$) and density dependence of nuclear symmetry energy (e.g. its slope $L$, curvature $K_{\rm{sym}}$, and skewness $J_{\rm{sym}} $around $\rho_0$).
\item The necessary spin parameter $\chi$ of GW190814's minor component of mass $m_2=(2.50-2.67)$M$_{\odot}$ \cite{abbottGW190814GravitationalWaves2020a} as a candidate of the most massive and fastest spinning NS \cite{Most:2020bba,zhang2020gw190814,Zhou:2020xan}.
\end{enumerate}
By applying the revised redshift $z$ data in a Gaussian likelihood function to the maximum mass configuration in the NS mass-radius sequence for a given EOS in our Bayesian analyses, we can ensure that all lower mass NSs also satisfy the $z$ constraint as we shall demonstrate in more detail.
We found that the skewness $J_0$ characterizing the stiffness of high-density SNM decreases significantly compared to its fiducial value extracted earlier from analyzing other astrophysical observations and nuclear collective flow in relativistic heavy-ion collisions. On the other hand, the slope $L$, curvature $K_{\rm{sym}}$, and skewness $J_{\rm{sym}}$ all increase appreciably, indicating a stiffer nuclear symmetry energy, especially at supersaturation densities compared to their fiducial values from earlier analyses without considering any $z$ constraint. Moreover, because the maximum mass (M$_{\rm{TOV}}\equiv$M$_{\rm{max}}$) of static NSs supported by a given nuclear EOS is reduced by the $z$ data while rotations generally increase the maximum masses of stable pulsars by up to only about 20\% at their Kepler frequencies, the peak of the posterior probability distribution function (PDF) of the spin rate is found to further move towards the Kepler frequency once the revised $z$ data is considered, making the minor component $m_2$ of GW190814 less likely to be a stable most massive pulsar observed so far.

The rest of the paper is organized as follows. In the next section, we shall first outline the meta-model EOS for NS matter and then discuss where we expect to see the effects of the revised redshift $z$ from GS 1826-24 before presenting the Bayesian framework we use. In section \ref{Results}, we present and discuss our results. Finally, a summary is given.

\section{Theoretical framework}
Our work is carried out within a comprehensive Bayesian framework using a meta-model nuclear EOS.
We adopt the so-called minimum NS model assuming NSs are made of neutrons, protons, electrons and muons (usually referred to as the npe$\mu$ matter) without considering the possible exotic phases, e.g., hyperons or hadron-quark phase transition inside NSs. Such an approach has been widely used by many groups in the literature.
We have previously used it to infer the nuclear EOS parameters by solving the NS inverse structure problem by brute-force and Bayesian statistical inference as well as forward-modeling of NS properties, see, e.g., Refs. \cite{Li:2019xxz,Li:2021thg} for reviews. To facilitate the presentation and discussions of our results, we recall here a few terminologies and outline several key aspects of our approach. We refer the reader to our earlier publications (e.g.
\cite{zhang2018combined,Zhang:2019fog,Zhang:2021xdt,xie2019bayesian,xie2020bayesian,xie2021bayesian})
for more details.

\subsection{Explicitly isospin-dependent meta-model EOS for neutron stars}
\label{nseos}
According to virtually all existing nuclear many-body theories, at the hadronic physics level the most fundamental quantity for computing the EOS of neutron-rich matter at nucleon density $\rho=\rho_n+\rho_p$ and isospin asymmetry $\delta\equiv (\rho_n-\rho_p)/\rho$ is the mean energy per nucleon $E(\rho ,\delta )$. It is a parabolic function of $\delta$ \cite{bombaci1991asymmetric}
\begin{equation}\label{eos}
E(\rho,\delta)=E_0(\rho)+E_{\rm{sym}}(\rho)\cdot \delta ^{2} +\mathcal{O}(\delta^4)
\end{equation}
where $E_0(\rho)$ is the EOS of SNM and $E_{\rm{sym}}(\rho)$ is nuclear symmetry energy at density $\rho$.
The pressure in NSs can be expressed as
\begin{equation}\label{pressure}
  P(\rho, \delta) = \rho^2 \frac{d\epsilon(\rho,\delta)/\rho}{d\rho},
\end{equation}
where $\epsilon(\rho, \delta) = \epsilon_n(\rho, \delta) + \epsilon_l(\rho, \delta)$ is the energy density of NS matter. Here, $\epsilon_n(\rho, \delta)$ and $\epsilon_l(\rho, \delta)$ are the energy densities of nucleons and leptons, respectively. The energy density of leptons, $\epsilon_l(\rho, \delta)$, is determined by utilizing the noninteracting Fermi gas model \cite{oppenheimer1939massive}. On the other hand, the energy density of nucleons, $\epsilon_n(\rho, \delta)$, is linked to the $E(\rho, \delta)$, and the average mass $M_N$ of nucleons via
\begin{equation}\label{lepton-density}
  \epsilon_n(\rho, \delta)=\rho [E(\rho,\delta)+M_N].
\end{equation}
The fractions of various particles (consequently the density profile of isospin asymmetry $\delta(\rho)$) can be obtained by solving the $\beta$-equilibrium condition $\mu_n-\mu_p=\mu_e=\mu_\mu\approx4\delta E_{\rm{sym}}(\rho)$ and the charge neutrality requirement $\rho_p=\rho_e+\rho_\mu$. Here the chemical potential $\mu_i$ for a particle $i$ is calculated from the energy density via
$\mu_i=\partial\epsilon(\rho,\delta)/\partial\rho_i.
$
With the above inputs, the density profile of isospin asymmetry $\delta(\rho)$ can be readily obtained.
The pressure in Eq. (\ref{pressure}) then becomes
barotropic, i.e. $P(\rho)$ depending on the density only. Similarly, the energy density $\varepsilon(\rho, \delta(\rho))$ becomes $\varepsilon(\rho)$ (i.e. also barotropic).
The core EOS described above is then connected with the crust EOS at a transition density consistently determined by examining when the core EOS becomes thermodynamically unstable \cite{Lattimer:2006xb,kubis2007nuclear,Xu:2009vi}. We adopt the NV EOS\cite{negele1973neutron} for the inner crust and the BPS EOS\cite{baym1971ground} for the outer crust. Finally, the complete NS EOS in the form of pressure versus energy density $P(\varepsilon)$ is used in solving the Tolman-Oppenheimer-Vokolff (TOV) equations
\cite{tolman1934effect,oppenheimer1939massive}.

In forward-modelings of NS EOS, given an energy density function $E(\rho, \delta)$ expressed by Eq. (\ref{eos}), it can be further expanded
by writing $E_0(\rho)$ and $E_{\rm{sym}}(\rho)$
as Taylor series around the saturation density $\rho_0$ of SNM as
\begin{eqnarray}\label{E0para}
  E_{0}(\rho)&=&E_0(\rho_0)+\frac{K_0}{2}(\frac{\rho-\rho_0}{3\rho_0})^2+\frac{J_0}{6}(\frac{\rho-\rho_0}{3\rho_0})^3,\\
  E_{\rm{sym}}(\rho)&=&E_{\rm{sym}}(\rho_0)+L(\frac{\rho-\rho_0}{3\rho_0})+\frac{K_{\rm{sym}}}{2}(\frac{\rho-\rho_0}{3\rho_0})^2
  +\frac{J_{\rm{sym}}}{6}(\frac{\rho-\rho_0}{3\rho_0})^3\label{Esympara},
\end{eqnarray}
where $E_0(\rho_0)$=-16 MeV at $\rho_0=0.16/\rm{fm}^3$. The coefficients $K_0=9\rho_0^2[\partial^2 E_0(\rho)/\partial\rho^2]|_{\rho=\rho_0}$, and $J_0=27\rho_0^3[\partial^3 E_0(\rho)/\partial\rho^3]|_{\rho=\rho_0}$ are the SNM incompressibility and skewness, respectively. The $E_{\rm{sym}}(\rho_0)$, $L=3\rho_0[\partial E_{\rm{sym}}(\rho)/\partial\rho]|_{\rho=\rho_0}$, $K_{\rm{sym}}=9\rho_0^2[\partial^2 E_{\rm{sym}}(\rho)/\partial\rho^2]|_{\rho=\rho_0}$ and $J_{\rm{sym}}=27\rho_0^3[\partial^3 E_{\rm{sym}}(\rho)/\partial\rho^3]|_{\rho=\rho_0}$ represent the magnitude, slope, curvature and skewness of nuclear symmetry energy at $\rho_0$, respectively.
We emphasize here that while the above coefficients are all defined at $\rho_0$, the high-order derivatives characterize the high-density behaviors of
$E_0(\rho)$ and $E_{\rm{sym}}(\rho)$. For instance, the skewness parameters $J_0$ and $J_{\rm{sym}}$ characterize the stiffness of
$E_0(\rho)$ and $E_{\rm{sym}}(\rho)$ around $(3-4)\rho_0$ \cite{Xie:2020kta},  and $K_{\rm{sym}}$ characterizes the stiffness of $E_{\rm{sym}}(\rho)$ around $(2-3)\rho_0$ which is most important for determining the radii of canonical neutron stars \cite{Richter:2023zec}.

\begin{figure}[h]
\begin{center}
  \resizebox{0.65\textwidth}{!}{
  \includegraphics[width=5cm,height=4cm]{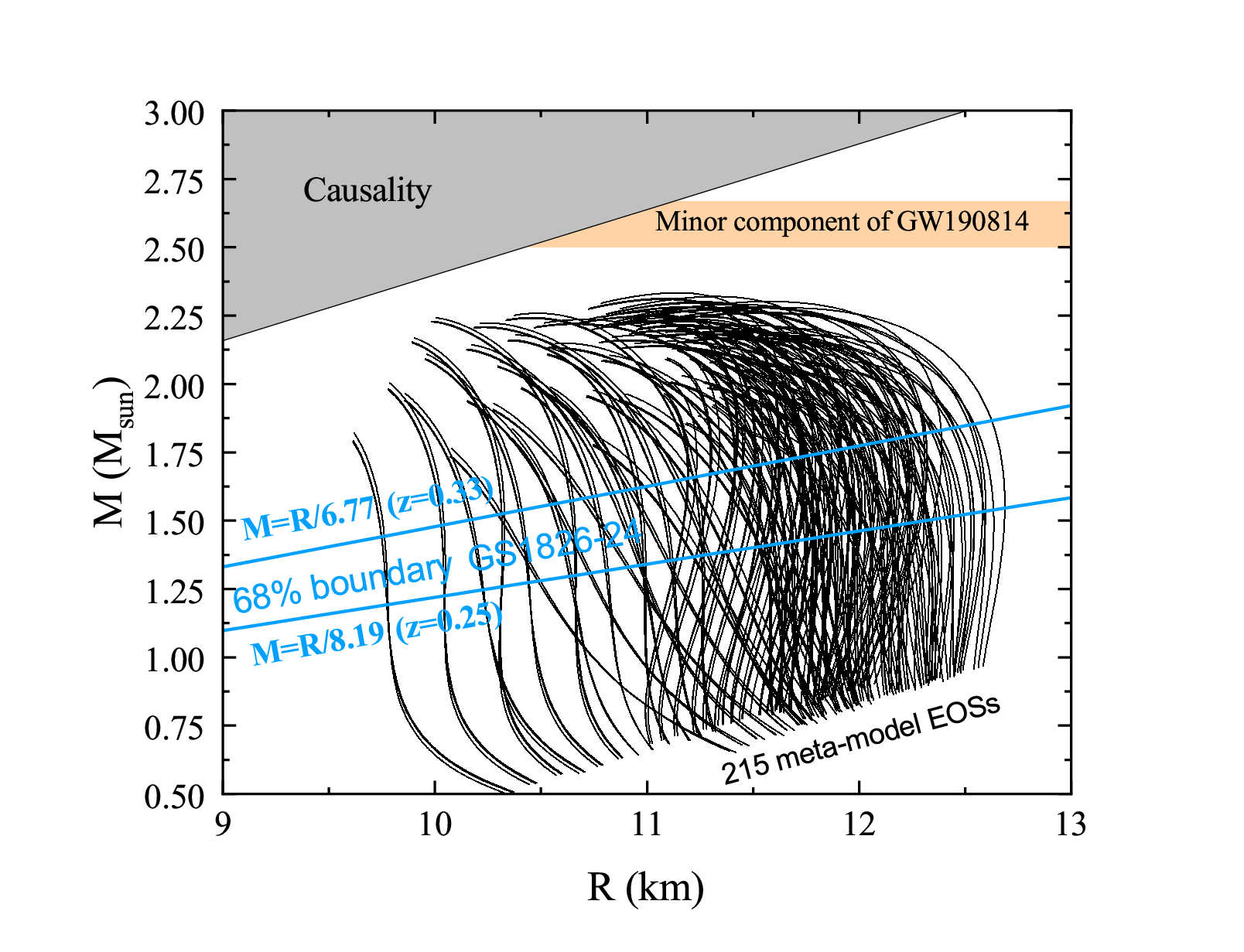}
}
  \caption{(color online) Mass-radius sequences from using 215 examples of meta-model EOSs in comparison with the 68\% upper (${\rm M}={\rm R}/6.77$ and lower (${\rm M}={\rm R}/8.19$) boundaries from the redshift of GS 1826-24. The horizontal band is the mass $m_2=2.59\pm 0.05$M$_{\odot}$ of GW190814's minor component.}\label{MR1plot}
\end{center}
\end{figure}

It is also worth noting that the above expressions and associated parameters have dual meanings.
They are Taylor expansions for known energy density functionals. In Bayesian analyses, however, they are just parameterizations with their six parameters to be inferred from the observational/experimental data instead of from actually expanding some known energy density functionals \cite{Cai:2021ucx}.
The issue of convergence associated with Taylor expansions does not exist in Bayesian analyses. In essence, the last parameters, i.e. $J_0$ and $J_{\rm{sym}}$, carry all information about the high-density EOS of neutron-rich matter. Of course, asymptotically they become the coefficients of the Taylor expansions as $\rho\rightarrow \rho_0$.
In Bayesian analyses, once the EOS parameters are randomly generated within their prior ranges in each Markov-Chain Monte Carlo (MCMC) step, the NS EOS can be constructed as described above. The posterior PDFs of the EOS parameters inferred will then be used to construct the experimentally/observationally preferred EOS according to the parameterizations of Eqs. (\ref{E0para}) and (\ref{Esympara}).
By varying or randomly generating the six EOS parameters, the NS EOS model described above becomes a meta-model, i.e. model of models. It can effectively mimic essentially all existing EOS models based on various microscopic and/or phenomenological nuclear many-body theories \cite{Li:2019xxz,Li:2021thg}.

\begin{figure}[h]
\begin{center}
   \resizebox{0.495\textwidth}{!}{
  \includegraphics[width=12cm,height=9.5cm]{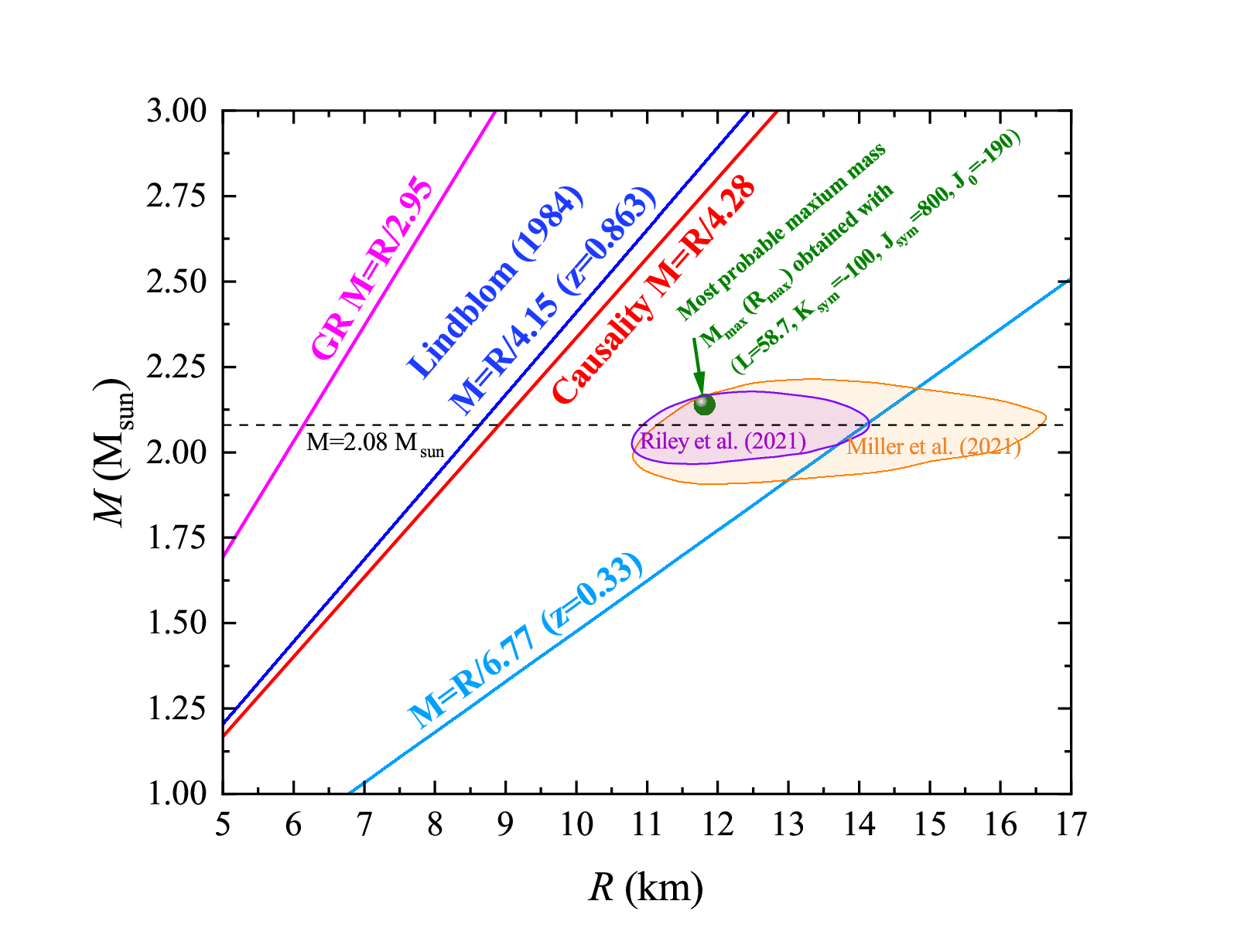}
  }
     \resizebox{0.495\textwidth}{!}{
   \includegraphics[width=12cm,height=9.5cm]{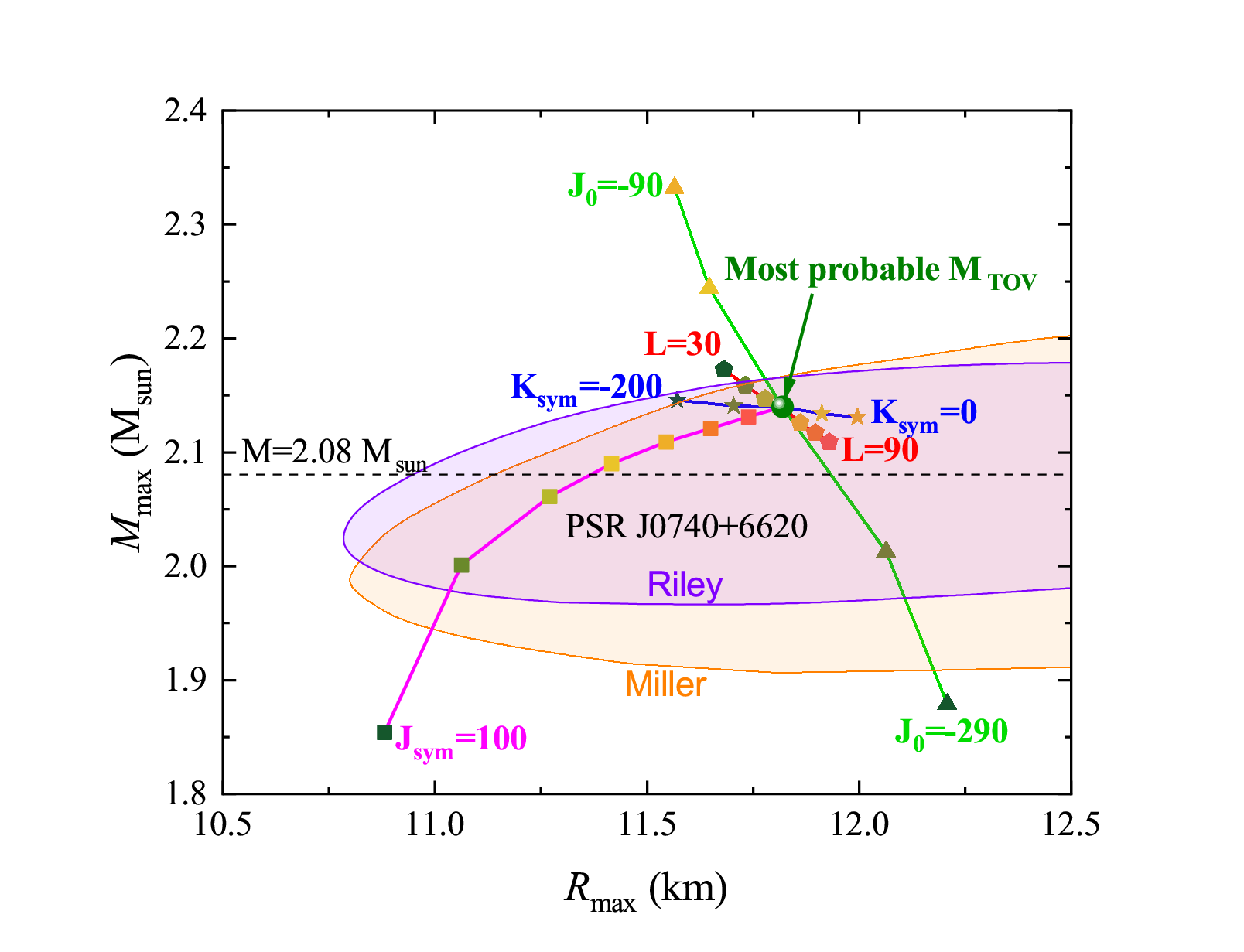}
   }
  \caption{(color online) Left: Constraints on the NS mass (M)-radius (R) plane from general relativity (GR) (Schwarzschild radius: $R\geq2GM/c^2$), causality ($R\geq2.9GM/c^2$), see Ref. \cite{Lattimer:2006xb} and references therein, earlier prediction for the upper bound of NS redshift z ($z<0.863$) by Lindblom \cite{Lindblom84}, and the new upper limit of z ($z<0.33$) from re-analyzing the light curve of NS in GS-1826-24 \cite{zhouMassMeasurementsShow2023a}. The shadowed ranges correspond to the 68\% mass-radius contours from two independent analyses using similar techniques of the same PSR J0740+6620 data from NICER \cite{miller2021radius,Riley:2021pdl}. The green dot in both panels is the most probable NS maximum mass M$_{\rm max}=2.14$ M$_{\odot}$ (M$_{\rm TOV}$) and corresponding radius R$_{\rm max}$ obtained with the parameter set: $L=58.7$ MeV, $K_{\rm sym}=-100$ MeV, $J_{\rm sym}=800$ MeV, and $J_{0}=-190$ MeV based on an earlier Bayesian analysis of astrophysical data \cite{xie2020bayesian}. Right: Dependence of M$_{\rm max}$ and R$_{\rm max}$ on the EOS parameters $L$, $K_{\rm sym}$, $J_{\rm sym}$, and $J_{0}$ in their respective uncertainty ranges of $30\leq L\leq90$ in steps of 10 MeV, $-200\leq K_{\rm sym}\leq0$ in steps of 50 MeV, $100\leq J_{\rm sym}\leq800$ in steps of 100 MeV, and $-290\leq J_0\leq-90$ in steps of 50 MeV.}\label{MmaxRmax}
\end{center}
\end{figure}
\subsection{Where do we expect to see effects of the revised redshift of X-ray burster GS-1826-24?}
Since the NS redshift or compactness measures its mass/radius ratio, it was generally considered a weak constraint on the EOS unless either the mass or radius is known from other sources \cite{Li:2005sr}. As an illustration, shown in Fig. \ref{MR1plot} are the mass-radius sequences from solving the TOV equations using 215 examples of meta-model EOSs in comparison with the 68\% upper (${\rm M}={\rm R}/6.77$) and lower (${\rm M}={\rm R}/8.19$) boundaries/contours from the redshift $z$ of GS 1826-24. These EOSs are generated by varying the EOS parameters within their current uncertainty ranges. The 68\% boundaries of the redshift $z$ from GS 1826-24 do not exclude any EOS as all currently acceptable EOSs pass through them. However, its high-$z$ tail may exclude the maximum mass (M$_{\rm TOV}$) configurations predicted by some of the EOSs considered. Moreover, it is seen that if the M$_{\rm TOV}$ configuration on the mass-radius sequence of a given EOS is compatible with the redshift observation, all lower-mass configurations from the same EOS can then automatically do so. Since the high-$z$ tail from GS 1826-24 represents small probabilities, a Bayesian statistical analysis is naturally useful for evaluating its impact on the nuclear EOS. As we shall discuss in detail, this can be done by imposing a $z$ component in the total likelihood function at the M$_{\rm TOV}$ configuration. It will then limit statistically the M$_{\rm TOV}$ and the posterior PDFs of the underlying EOS parameters.

To put the above expectation and its importance on a more solid physics background, shown in the left panel of Fig. \ref{MmaxRmax} with a green dot is the most probable M$_{\rm TOV}$ obtained with the most probable EOS parameters: $L=58.7$ MeV, $K_{\rm sym}=-100$ MeV, $J_{\rm sym}=800$ MeV, and $J_{0}=-190$ MeV based on an earlier Bayesian analysis \cite{xie2020bayesian} of astrophysical data including those for PSR J0740+6620 from NICER \cite{miller2021radius,Riley:2021pdl} under constraints from nuclear experiments and theories \cite{Li:2021thg}. The PSR J0740+6620 having an updated mass of $2.08\pm 0.07~M_{\odot}$ \cite{Fonseca:2021wxt} is among the most massive NSs discovered so far. Its radius of $13.7^{+2.6}_{-1.5}$ km (68\%) from Miller {\it et al.} \cite{miller2021radius} or $12.39_{-0.98}^{+1.30}$\,km from Riley {\it et al.} \cite{Riley:2021pdl} was inferred from two independent analyses of the same X-ray data taken by the \textit{Neutron Star Interior Composition Explorer} (NICER) and the X-ray Multi-Mirror (XMM-Newton) observatory. It is seen that the mass-radius contour for PSR J0740+6620 from Riley {\it et al.} is mostly above the 68\% upper boundary of the redshift $z$ from GS 1826-24. Nevertheless, assuming both have Gaussian distributions, their tails overlap. On the other hand, the 68\% upper boundary of the redshift $z$ from GS 1826-24 passes through the mass-radius contour from Miller {\it et al.}. It would thus be exciting to investigate to what extent the high-$z$ tail of GS 1826-24 may affect the EOS parameters compared to our prior knowledge from analyzing earlier astrophysical data including those from NICER.

It is also seen that the most probable M$_{\rm TOV}$ (the green dot at $2.14$ M$_{\odot}$) is significantly above the 68\% upper boundary of the redshift $z$ from GS 1826-24. However, the $1\sigma$ uncertainties of the underlying EOS parameters, especially $J_0$ and $J_{\rm sym}$ are still considerable. Consequently, as shown in the right panel, the M$_{\rm TOV}$ varies significantly with the EOS parameters within their $1\sigma$ uncertainties. In these calculations, for example, in analyzing the impact of the parameter $J_0$ on the M$_{\rm TOV}$ and the corresponding radius R$_{\rm max}$ of NSs, we systematically vary the values of $J_0$ from -290 to -90 MeV while keeping all other parameters fixed at their most probable values given above. Our analysis reveals that the parameter $J_0$ exhibits greater influences on the M$_{\rm TOV}$ than the radius R$_{\rm max}$ when it is varied from -290 to -90 MeV, consistent with our earlier findings \cite{zhang2018combined,Zhang:2019fog,Zhang:2021xdt,xie2019bayesian,xie2020bayesian,xie2021bayesian}. Conversely, the parameters $J_{\mathrm{sym}}$ and $K_{\mathrm{sym}}$ demonstrate more pronounced influences on the radius R$_{\rm max}$ than on the M$_{\rm TOV}$ when they are varied from 100 to 800 MeV and from -200 to 0 MeV, respectively. While the parameter $L$ is relatively well constrained to be between 30 and 90 MeV, it is still obvious that its variation affects significantly the R$_{\rm max}$ but has little effect on the M$_{\rm TOV}$. Interestingly, most of the results from varying the key EOS parameters fall within the 68\% error band of PSR J0740+6620 observed by NICER \cite{miller2021radius,Riley:2021pdl}. Among all EOS parameters considered, $J_0$ has the strongest influence on M$_{\rm TOV}$ while $J_{\rm sym}$ has the biggest impact on R$_{\rm max}$.

It is also interesting to note that various efforts within other models have put the M$_{\rm TOV}$ in the range of about 2.1 to 2.4 M$_{\odot}$ (see Fig. 5.1 of Ref. \cite{Li:2019xxz} for a summary) consistent with the results shown in the right panel of Fig. \ref{MmaxRmax}. For example, a very recent comprehensive
analysis \cite{Fan:2023spm} extracted an M$_{\rm TOV}=2.25^{+0.08}_{-0.07}$ M$_{\odot}$
by combining results from two approaches: (1)
modeling the mass function of 136 NSs with a sharp cut-off at 2.28 M$_{\odot}$, and (2) Bayesian inference of the EOS from the multi-messenger data of NSs under the constraints of predictions from the chiral effective field theory at low densities and perturbative QCD at extremely high densities.
In short, our current knowledge on M$_{\rm TOV}$ is model dependent albeit within a relatively small range. Fortunately, the new redshift data from GS 1826-24 provides us with an additional constraint on the M$_{\rm TOV}$.

Another impact of the constraint on M$_{\rm TOV}$ due to the high-$z$ tail of GS 1826-24 is on the nature of GW190814's secondary component $m_2$.
Whether the $m_2$ is the lightest black hole or the most massive NS remains a challenging question that is still under hot debate, see, e.g. Ref. \cite{Sedrakian:2022ata} for a recent review.
Among many possible mechanisms for the $m_2$ to be a NS, the simplest one relies on the additional rotational support besides the nuclear pressure, see, e.g. Refs. \cite{Most:2020bba,zhang2020gw190814,Biswas:2020xna,Li:2020ias,Sedrakian:2020kbi,Demircik:2020jkc,Dexheimer:2020rlp,Khadkikar:2021yrj,Tsokaros:2020hli,Riahi:2020rkg,Zhou:2020xan}. In these models, the required minimum spin frequency depends on the M$_{\rm TOV}$. There is a universal relation \cite{breuMaximumMassMoment2016,musolino2024maximum}
\begin{equation}
M_{\rm max}^{\rm rot} = M_{\rm TOV} \left[1+a_2\left(\frac{\chi}{\chi_{\rm kep}}\right)^2 +a_4\left(\frac{\chi}{\chi_{\rm kep}}\right)^4\right]\,
\label{breu and rezolla:universal relation}
\end{equation}
between the maximum mass $M_{\rm max}^{\rm rot}$ of a rotating NS (pulsar), M$_{\rm TOV}$ as well as its dimensionless spin magnitude $\chi$ (relative to its maximum value $\chi_{\rm kep}$ at the Kepler frequency, i.e. the mass-shedding limit). Following Ref. \cite{musolino2024maximum}, the parameters $a_2$ and $a_4$ are selected according to the scheme: if M$_{\rm TOV}$ is greater than twice the solar mass (M$_{\rm TOV}\ge 2.0$ M$_{\odot}$), $a_2$=0.16 and $a_4$=$\cal R$-$(a_2+1)$ with $\cal R$=1.24; $a_2$=0.17 with $\cal R$=1.244 if M$_{\rm TOV}\ge 2.2$ M$_{\odot}$; and $a_2$=0.16 with $\cal R$=1.248 if M$_{\rm TOV}\ge 2.35$ M$_{\odot}$, respectively.
The horizontal band in Fig. \ref{MR1plot}
is the mass $m_2=2.50-2.67$ of the minor component of GW190814 at 68\% confidence level.
To rotationally raise the various $M_{\rm TOV}$ values to be combative with the $m_2$ requires different
spins. Constraints on the M$_{\rm TOV}$ from the high-$z$ tail of GS 1826-24 will affect the necessary spin to rotationally support the $m_2$ of GW190814. The resulting posterior PDF of $\chi$ will thus provide additional knowledge useful for better understanding the nature of GW190814's minor component $m_2$.

\subsection{Bayesian inference approach}\label{bayes}
For completeness, we recall that the Bayesian theorem reads
\begin{equation}\label{Bay1}
P({\cal M}|D) = \frac{P(D|{\cal M}) P({\cal M})}{\int P(D|{\cal M}) P({\cal M})d\cal M}.
\end{equation}
Here, $P({\cal M}|D)$ represents the posterior probability of the model $\cal M$ given the dataset $D$, which is the focal point of our inquiry. Meanwhile, $P(D|{\cal M})$ denotes the likelihood function or the conditional probability that a given theoretical model $\cal M$ accurately predicts the data $D$. Additionally, $P({\cal M})$ signifies the prior probability of the model $\cal M$ before encountering the given data. The denominator in Eq. (\ref{Bay1}) represents a normalization constant that can be used to discern which model among many candidates can provide a more rational explanation for experimental data.
\begin{table*}[htbp]
\centering
\caption{Prior ranges of the six EOS parameters (in units of MeV) and the reduced spin parameter $\chi/\chi_{\rm kep}$.}\label{tab-prior}
 \begin{tabular}{lccccccc}
  \hline\hline
   Parameters&~~~~Lower limit  &~~~~Upper limit \\
    \hline\\
  \vspace{0.2cm}
$K_0$ & 220 & 260 \\
$J_0$ & -800 & 400 \\
$K_{\mathrm{sym}}$  & -400 & 100 \\
$J_{\mathrm{sym}}$ & -200 & 800 \\
$L$ & 30 & 90 \\
$E_{\mathrm{sym}}(\rho_0)$ & 28.5 & 34.9 \\
$\chi/\chi_{\rm kep}$ &0.2& 1.0\\
 \hline
 \end{tabular}
\end{table*}

\begin{table*}[htbp]
\centering
\caption{Data for the NS radius, maximum mass and redshift used in the present work.}\label{tab-data}
 \begin{tabular}{lccccccc}
  \hline\hline
   Mass($\mathrm{M}_{\odot}$)&Radius $R$ (km)  &~~~~Source and Reference \\
    \hline\hline\\
  \vspace{0.2cm}
 1.4 & 11.9$^{+1.4}_{-1.4}$(90\% CFL)&GW170817\cite{abbott2018gw170817} \\
1.4 &10.8$^{+2.1}_{-1.6}$ (90\% CFL)&GW170817 \cite{de2018tidal} \\
1.4  & 11.7$^{+1.1}_{-1.1}$ (90\% CFL)&QLMXBs \cite{lattimer2014constraints} \\
$1.34_{-0.16}^{+0.15}$ &$12.71_{-1.19}^{+1.14}$ (68\% CFL)&PSR J0030+0451 \cite{riley2019nicer} \\
$1.44_{-0.14}^{+0.15}$ &$13.0_{-1.0}^{+1.2}$ (68\% CFL)&PSR J0030+0451 \cite{fonseca2021refined} \\
$2.08_{-0.07}^{+0.07}$ &$13.7_{-1.5}^{+2.6}$ (68\% CFL)&PSR J0740+6620 \cite{fonseca2021refined} \\
  \hline\hline
  &Redshift &Source and Reference \\
    \hline\\
  \vspace{0.2cm}
 redshift-1 &$0.29_{-0.04}^{+0.04}$ (68\% CFL)  &GS 1826-24\cite{zhouMassMeasurementsShow2023a}\\
 redshift-2 &$0.29_{-0.034}^{+0.034}$,$0.25_{-0.0065}^{+0.0065}$  (68\% CFL)  &GS 1826-24\cite{zhouMassMeasurementsShow2023a}\\
  \hline\hline
  Maximum Mass &Source and Reference \\
    \hline\\
  \vspace{0.2cm}
 $2.59_{-0.05}^{+0.05}$ (68\% CFL)  &GW190814\cite{abbottGW190814GravitationalWaves2020a}\\
 \hline
 \end{tabular}
\end{table*}
\begin{figure}[h]
\begin{center}
  \resizebox{0.65\textwidth}{!}{
  \includegraphics[width=5cm,height=4cm]{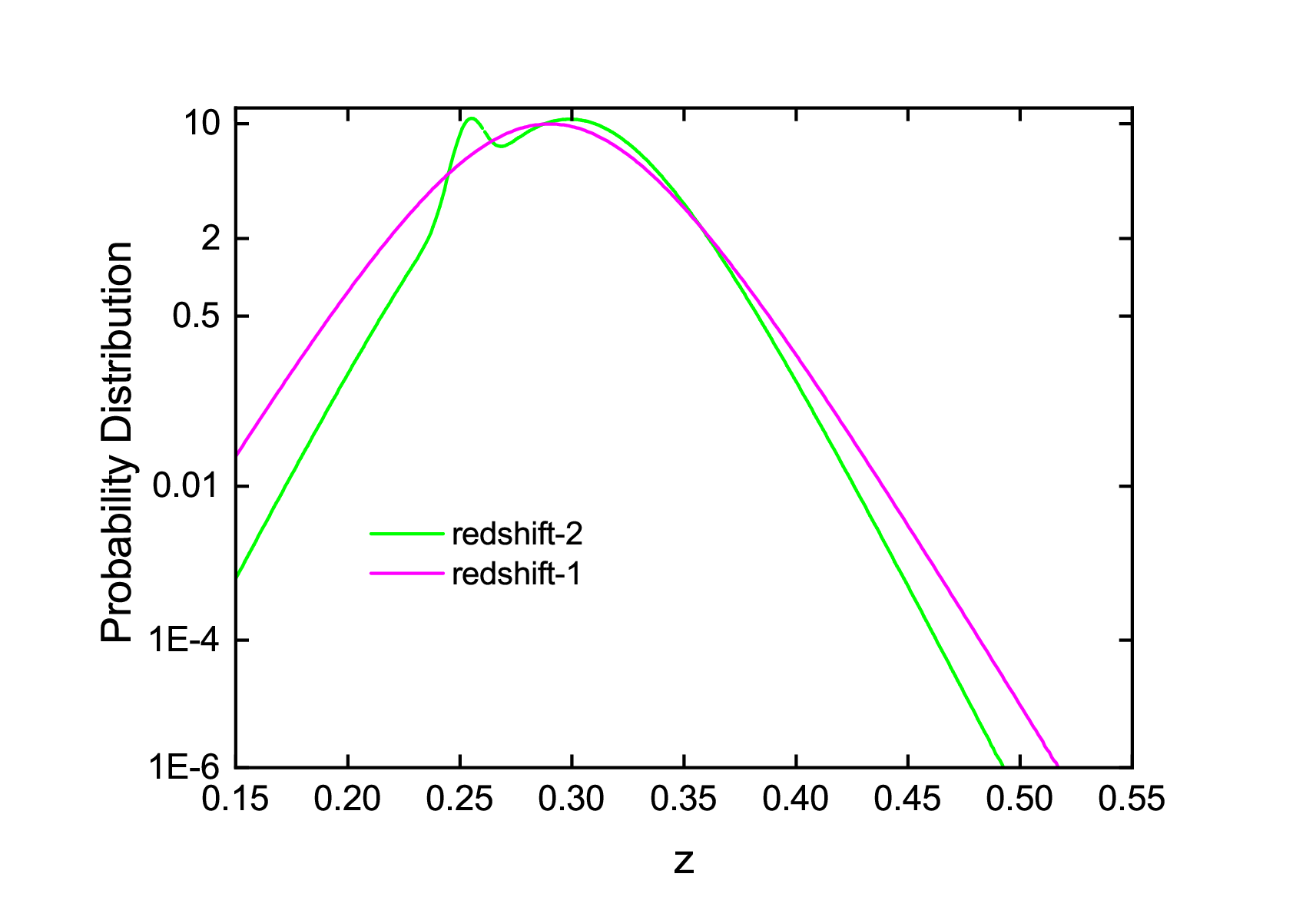}
}
  \caption{(color online) The gravitational redshift-z distributions extracted from Ref. \cite{zhouMassMeasurementsShow2023a}. The ``redshift-1" is a Gaussian function based on the information about the 95\% confidence contour,  while the ``redshift-2" is a weighted sum of two Gaussians based on the information about the 68\% confidence contours in the distance-redshift plane given in Ref. \cite{zhouMassMeasurementsShow2023a}.}\label{zPDF}
\end{center}
\end{figure}
We randomly sample uniformly the EOS and pulsar spin parameters within their prior ranges specified by their minimum and maximum values shown in Table \ref{tab-prior}. Through substitution of the six randomly sampled EOS parameters, $p_{i=1,2\cdots 6}$, into equations (\ref{pressure}), (\ref{E0para}), and (\ref{Esympara}), we construct the EOS for NSs at $\beta-$equilibrium. Subsequently, the NS mass-radius sequences are determined by solving the TOV equations. The resultant theoretical radius $R_{\mathrm{th},j}$ is then utilized to assess the likelihood of the selected EOS parameter set to reproduce the observed radius $R_{\mathrm{obs},j}$, where $j$ ranges from 1 to 6, as presented in the data set $D(R_{1,2,\cdots 6})$ in Table \ref{tab-data}. This radius likelihood calculation is expressed as follows:
\begin{equation}\label{Likelihood-R}
  P_\mathrm{R}[D(R_{1,2,\cdots 6})|{\cal M}(p_{1,2,\cdots 6})]=\prod_{j=1}^{6}\frac{1}{\sqrt{2\pi}\sigma_{\mathrm{obs},j}}\exp[-\frac{(R_{\mathrm{th},j}-R_{\mathrm{obs},j})^{2}}{2\sigma_{\mathrm{obs},j}^{2}}],
\end{equation}
where $\sigma_{\mathrm{obs},j}$ represents the $1\sigma$ error bar associated with the observation $j$. For the data having different upper and lower 68\% confidence boundaries ($\sigma)$, we use them separately for $R$ smaller and larger than $R_{\mathrm{obs}}$ to mimic a finally asymmetric (non) Gaussian distribution.
Furthermore, when multiple results exist from different analyses of observations of the same source, such as GW170817 and PSR J0030+0451, we consider them as equally probable. Numerically, we take their statistical average by randomly selecting with an equal weight one of the results to calculate the likelihood function in each MCMC step.

After determining the mass and radius of a NS, its surface gravitational redshift can be computed using the expression $z = \left( 1 - 2 M/R \right)^{-1/2} - 1$. The corresponding redshift likelihood function is formulated as:
\begin{equation}\label{Likelihood-z}
P(z,z_{\mathrm{obs}},\sigma_{\mathrm{z}}) = \frac{1}{\sqrt{2\pi}\sigma_{\mathrm{z}}}{\exp[-\frac{(z-z_{\mathrm{obs}})^{2}}{2\sigma_{\mathrm{z}}^{2}}]}.
\end{equation}
It is necessary to discuss here in more detail what redshift information we can get from Ref. \cite{zhouMassMeasurementsShow2023a} and how we use it in determining the z-likelihood function $P(z,z_{\mathrm{obs}},\sigma_{\mathrm{z}})$ in each MCMC step. Ref. \cite{zhouMassMeasurementsShow2023a} published the probability contours at 95\% and 68\% confidence level, respectively, in the distance-redshift plane. The 95\% contour centers around $z_{0} = 0.29$ with $\sigma_{\mathrm{z_0}} = 0.04$ ($1\sigma$ calculated from the $1.96\sigma$ on the 95\% line assuming the z-distribution is Gaussian), while there are two 68\% contours: a narrow one around $z_1 = 0.25$ with $\sigma_{\mathrm{z}_1} = 0.0065$ and a broad one around $z_2 = 0.29$ with $\sigma_{\mathrm{z}_2} = 0.034$, respectively. The difference of 0.006 between $\sigma_{\mathrm{z_0}}$ and $\sigma_{\mathrm{z}_2}$ is due to the non-Gaussian nature of the actual z-distribution that is not available to us. We treat it as an uncertainty of our analyses. Without the detailed distance-redshift joint distribution function and/or extra information about the distance, our extraction of the z-distribution inevitably involves some uncertainties. Nevertheless, we faithfully use the available data and try to mitigate the potential impact of the involved uncertainties on our conclusions by using two observational z-distributions (likelihood) functions. Our ``redshift-1" is just the $D_1(z)=P(z,z_0,\sigma_{\mathrm{z_0}})$ based on the 95\% confidence contour mentioned above. While our ``redshift-2" is non-Gaussian constructed from the weighted sum of two Gaussians, i.e., $D_2(z)=P(z,z_{1},\sigma_{\mathrm{z_1}})+\lambda P(z,z_{2},\sigma_{\mathrm{z_2}})$
with $\lambda=8.96$ by requiring the resulting $D_2(z)$ to have the same value at $z=z_1$ and $z=z_2$ to be consistent with the two 68\% contours given in Ref. \cite{zhouMassMeasurementsShow2023a}. Technically, this is simply done by comparing a random number generated between 0 and 1 with $1/(1+\lambda)\approx 10\%$ (which is the total probability $P_{\rm low}(z)$ for the narrower Gaussian peaked at z=0.25) to select one of the two Gaussians involved in $D_2(z)$ in each MCMC step as we are doing a statistical Bayesian analysis. Interestingly, we found numerically that our results are almost identical with  $P_{\rm low}(z)$=0.1 or 0.5 as what matters for this study is the high-z tail at the M$_{\rm TOV}$ configuration as we discussed earlier. Consequently, considering the narrow z-distribution around $z_1 = 0.25$ or not has practically no effect on our study as the high-z tail is overwhelmingly determined by the broader Gaussian distribution peaked at z=0.29. Since what really matters for the purposes of this work is the high-z tail, it is useful to compare the $D_1(z)$ and $D_2(z)$ especially at high-z values. Their log plots are shown in Fig.\ \ref{zPDF}. With the limited data available,
while our likelihood function has some uncertainties, the truth should be between the redshift-1 and redshift-2 and more closer to the latter.
It is seen that the redshift-2 z-distribution has a less extended high-z tail. Namely, it favours relatively smaller z-values for the M$_{\rm TOV}$ configuration compared to redshift-1. Thus, the redshift-2 is expected to lead to a softer SNM EOS and a stiffer symmetry energy compared to the redshift-1 to produce a smaller compactness. Comparing results from using the redshift-1 and redshift-2 will thus further demonstrate effects of the high-z tail of the gravitational redshift with respect to calculations with no limit on $z$ until the causality line. 

By substituting the randomly sampled spin parameter $\chi/\chi_{\rm kep}$ along with the M$_{\rm TOV}$ into equation (\ref{breu and rezolla:universal relation}), the pulsar maximum mass $M_{\rm max}^{\rm rot}$ is then derived. Subsequently, its likelihood function to reproduce the observed mass $m_2$ of GW190814 is computed from
\begin{equation}\label{Likelihood-M}
  P_\mathrm{M}[D(M^{\mathrm{max}})|{\cal M}(p_{1,2,\cdots 7})] = \frac{1}{\sqrt{2\pi}\sigma_{\mathrm{M}}}{\exp[-\frac{(M^{\mathrm{max}}_{\mathrm{th}}-M^{\mathrm{max}}_{\mathrm{obs}})^{2}}{2\sigma_{\mathrm{\mathrm{M}}}^{2}}]},
\end{equation}
where $\sigma_{\mathrm{M}} = 0.05$, and $M^{\mathrm{max}}_{\mathrm{obs}} = 2.59$, as listed in Table \ref{tab-data}.

Depending on the purpose of each calculation, the total likelihood function can be constructed differently by selecting and multiplying the individual likelihood components given above. For example, our base/default total likelihood function is given by
\begin{equation}\label{Likelihood}
  P(D|{\cal M})_{\rm base} = P_{\rm{filter}} \times P_{\rm{mass,max}} \times P_\mathrm{R}.
\end{equation}
Here, $P_{\rm{filter}}$ and $P_{\rm{mass,max}}$ denote that the generated EOSs must satisfy the following conditions: (i) The crust-core transition pressure remains positive; (ii) The thermodynamic stability condition, $dP/d\varepsilon\geq0$, holds at all densities; (iii) The causality condition is upheld at all densities; (iv) The generated NS EOS should be sufficiently stiff to support the currently observed maximum mass of NSs. Namely, the $P_{\rm{mass,max}}$ is a step function at the minimum M$^{\rm min}_{\rm TOV}$. In this study, we set
it at 1.97 M$_{\odot}$ as in the original analysis of GW170817 by the LIGO/VIRGO Collaborations \cite{abbott2017gw170817}.

To evaluate the effects of redshift on the EOS parameters, our default results will be compared to calculations obtained from using a total likelihood function $P(D|{\cal M})= P(D|{\cal M})_{\rm base} \times P_\mathrm{z}$. In the case of studying its effects on the spin of GW190814's minor, we will compare the default results with calculation using $P(D|{\cal M})= P(D|{\cal M})_{\rm base} \times P_\mathrm{z} \times P_\mathrm{M}$.

The Metropolis-Hastings algorithm is utilized in our MCMC process to generate posterior PDFs for the model parameters. These PDFs not only encompass the individual parameters but also enable the calculation of two-parameter correlations by integrating all other parameters via the marginal estimation approach. During the initial phase, known as the burn-in period, samples are discarded to ensure that the MCMC process begins from an equilibrium distribution, as recommended by Trotta (2017)\cite{trotta2017bayesian}. Specifically, a total of 40,000 burn-in steps are thrown away, with one million subsequent steps used for calculating the posterior PDFs in each case.

\begin{figure}[ht]
  \centering
   \resizebox{0.85\textwidth}{!}{
  \includegraphics[bb=10 300 650 830]{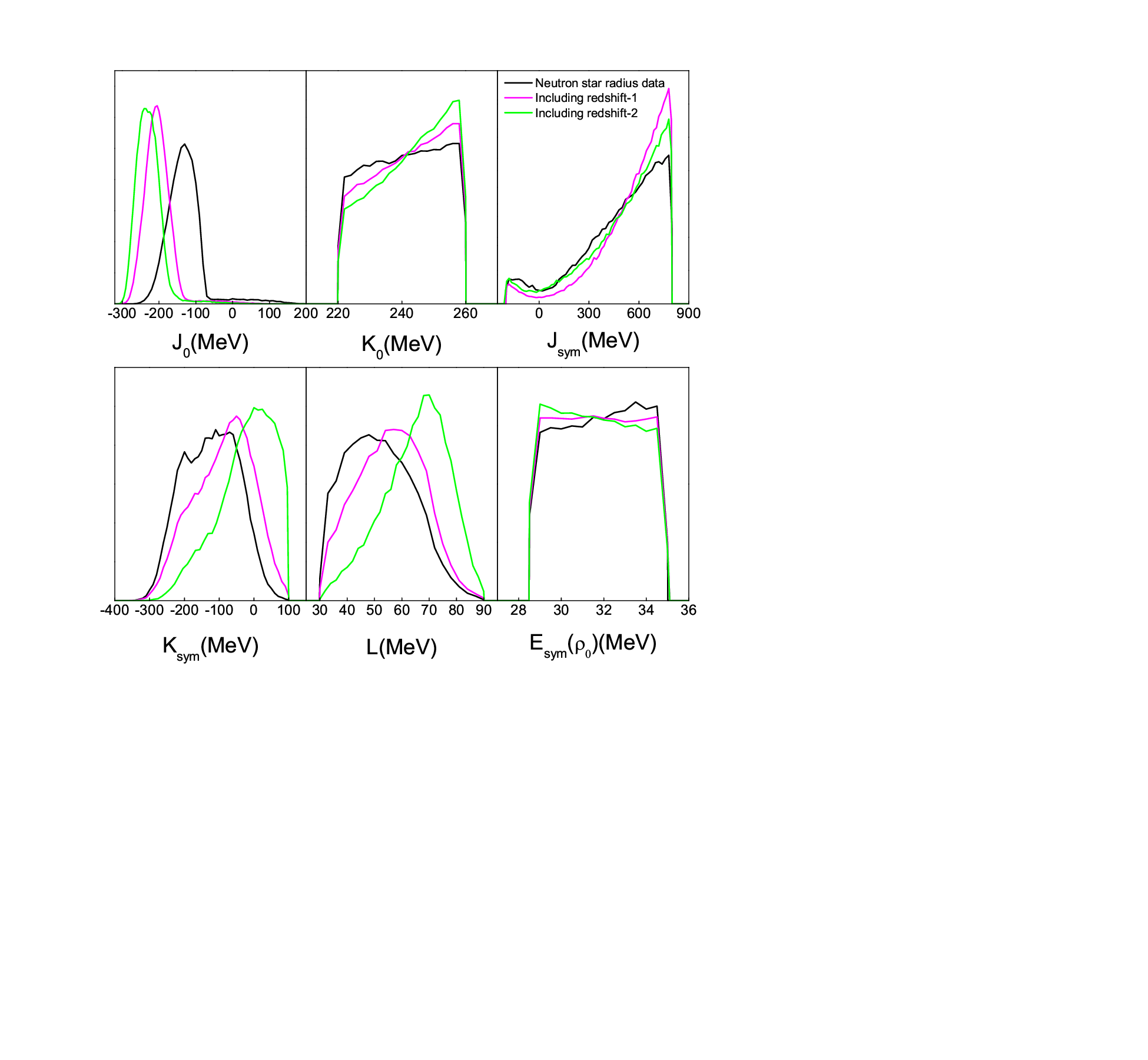}
  }
  \caption{(color online) Posterior probability distribution functions of the six EOS parameters inferred from the Bayesian analysis of combined NS radius dataset listed in Table \ref{tab-data} with (magenta and green) and without (black) considering the gravitational redshift of GS 1826-24. The results from the (non-)Gaussian form of the gravitational redshift distribution are represented by (green) magenta curve corresponding to the redshift-1 (redshift-2) shown in Fig.\ \ref{zPDF}.}\label{6para}
\end{figure}

\section{Results and Discussions}\label{Results}
\subsection{Impact on the EOS of supradense neutron-rich matter}
The posterior PDFs of the six EOS parameters from the radius data listed in Table \ref{tab-data} with and without considering the gravitational redshift constraint from GS 1826-24 are shown in Fig. \ref{6para}. It is seen that the skewness $J_0$ of SNM decreases significantly while the $L$, $K_{\rm sym}$ and $J_{\rm sym}$ characterizing the stiffness of nuclear symmetry all turn to increase. The incompressibility $K_0$ also shows an appreciable increase. While the magnitude of the symmetry energy $E_{\rm sym}$ at $\rho_0$ shows some slight decrease mostly due to its correlations with other parameters. These results are all understandable and interesting.
In particular, it is well known that the M$_{\rm TOV}$ is mostly determined by the stiffness of SNM at supersaturation densities characterized by $J_0$, see demonstrations in either forward-modelings in Ref. \cite{Zhang:2018bwq} and/or Bayesian analysis in Ref. \cite{xie2019bayesian}. While the NS radius in general is mostly determined by the stiffness of nuclear symmetry energy. A stiffer SNM EOS leads to
a higher M$_{\rm TOV}$ while a stiffer symmetry energy results in a larger radius R$_{\rm max}$.
The incompressibility $K_0$ has little effect on the M$_{\rm{TOV}}$ but its increase also leads to a larger radius at least for canonical NSs as shown earlier using the same meta-model NS EOS in Refs. \cite{Li:2020ass,Richter:2023zec}.
As discussed earlier and shown in Fig. \ref{MR1plot} and Fig. \ref{MmaxRmax}, the revised gravitational redshift of GS 1826-24 leads to a reduced NS compactness $\xi\equiv M/R$. At the maximum mass configuration, this requires either a reduction of M$_{\rm TOV}$ and/or an increase of the corresponding radius R$_{\rm max}$. Consequently, the high-density SNM EOS has to be softened while the symmetry energy at supersaturation densities stiffened.
As expected, this phenomenon becomes more pronounced when using the non-Gaussian distribution redshift-2 as it results in smaller values of gravitational redshift compared to the redshift-1. Corner plots of the pairwise correlation functions among the six-EOS parameters together with their PDFs for all three cases are shown in the Appendix.

\begin{table*}[htbp]
\centering
\caption{Most probable values and their corresponding credible intervals (68\% and 90\%) of the six EOS parameters inferred from the Bayesian analyses of NS radii with and without considering the gravitational redshift constraint from GS 1826-24. Here the phrase ``NS radius including redshift-1 (redshift-2)" means that the neutron-star radius data and the redshift-1 (redshift-2) z-distribution from the gravitational redshift data are used in the calculations. }\label{MP1}
\begin{tabular}{lccccccc}
  \hline\hline
  Parameters (MeV) &~~~~~~NS radius including redshift-1 &NS radius including redshift-2 &~~~~~~NS radius data only\\
  &68\%, 90\% &68\%, 90\% &68\%, 90\%\\
  \hline\hline\\
 $J_0:$ &$-205_{-35}^{+25},  -205_{-55}^{+45}$&$-240_{-20}^{+35}, -240_{-40}^{+55}$ &$-130_{-40}^{+30}, -130_{-70}^{+50}$\\
 $K_0:$ &$258_{-24}^{+2},  258_{-34}^{+2}$& $258_{-22}^{+2}, 258_{-32}^{+2}$ & $258_{-26}^{+2}, 258_{-34}^{+2}$  \\
 $J_{\mathrm{sym}}:$ &$800_{-275}^{+0},  800_{515}^{+0}$&$800_{-350}^{+0}, 800_{-635}^{+0}$ &$800_{-400}^{+0}, 800_{-660}^{+0}$ \\
 $K_{\mathrm{sym}}:$ &$-50_{-110}^{+60},  -50_{-180}^{+90}$&$24_{-84}^{+60}, 24_{-168}^{+72}$ &$-70_{-130}^{+30}, -70_{-180}^{+60}$ \\
 $L:$  &$57_{-15}^{+9}, 57_{-21}^{+15}$ &$70_{-14}^{+8}, 70_{-24}^{+14}$&$50_{-20}^{+10}, 50_{-20}^{+20}$\\
 $E_{\mathrm{sym}}(\rho_0):$ &$31.5_{-2.5}^{+1.5},  31.5_{-2.5}^{+3}$&$29.0_{-0}^{+4}, 29.0_{-0}^{+5.5}$ &$33.5_{-3}^{+1}, 33.5_{-4.5}^{+1}$ \\
  \hline
 \end{tabular}
\end{table*}
To characterize more quantitatively the effects of the revised gravitational redshift of GS 1826-24 on the EOS parameters, their most probable values and corresponding credible intervals (at 68\% and 90\% confidence levels) are listed in Table \ref{MP1}. Most noticeably, the $J_0$ decreases by about 58\% (85\%) while the $K_{\rm sym}$ increases by about 29\% (134\%) with the redshift-1 (redshift-2) z-distribution compared to their most probable values in the default calculations without considering the constraint from the gravitational redshift of GS 1826-24. The large changes (from 58\% to 85\% for $J_0$ and from 29\% to 134\% for $K_{\rm sym}$) due to changing the z-distribution from redshift-1 to redshift-2 strongly suggest that the gravitational redshift data are more sensitive to the variation of $R_{\rm max}$ than M$_{\rm TOV}$. Consequently, it is more revealing about the symmetry energy in the high-density region characterized by $K_{\rm sym}$ compared to the SNM EOS there characterized by $J_0$.
It is known that the radii and tidal deformations of all NSs observed so far have provided significant constraints on the $L$ and to a lesser extent on $K_{\rm sym}$, but not much on the $J_{\rm sym}$ as indicated by the peak of its PDF at the upper limit of its prior \cite{Li:2021thg}. The shift observed here in the PDF of $J_{\rm sym}$ is mostly due to its known anti-correlations with both $J_0$ and $K_{\rm sym}$ \cite{xie2019bayesian}, see the Appendix. Since $J_{\rm sym}$ determines mostly the behavior of symmetry energy above about $(3-4)\rho_0$, while the radii of canonical NSs are mostly determined by NS pressure at lower densities where $L$ and $K_{\rm sym}$ are more important, it is not surprising that the redshift data has the strongest impact on $K_{\rm sym}$ among all symmetry energy parameters although neither the mass nor radius (but their ratio) of the NS in GS 1826-24 is known.

\begin{figure}[ht]
  \centering
   \resizebox{0.45\textwidth}{!}{
  \includegraphics[bb=10 270 400 680]{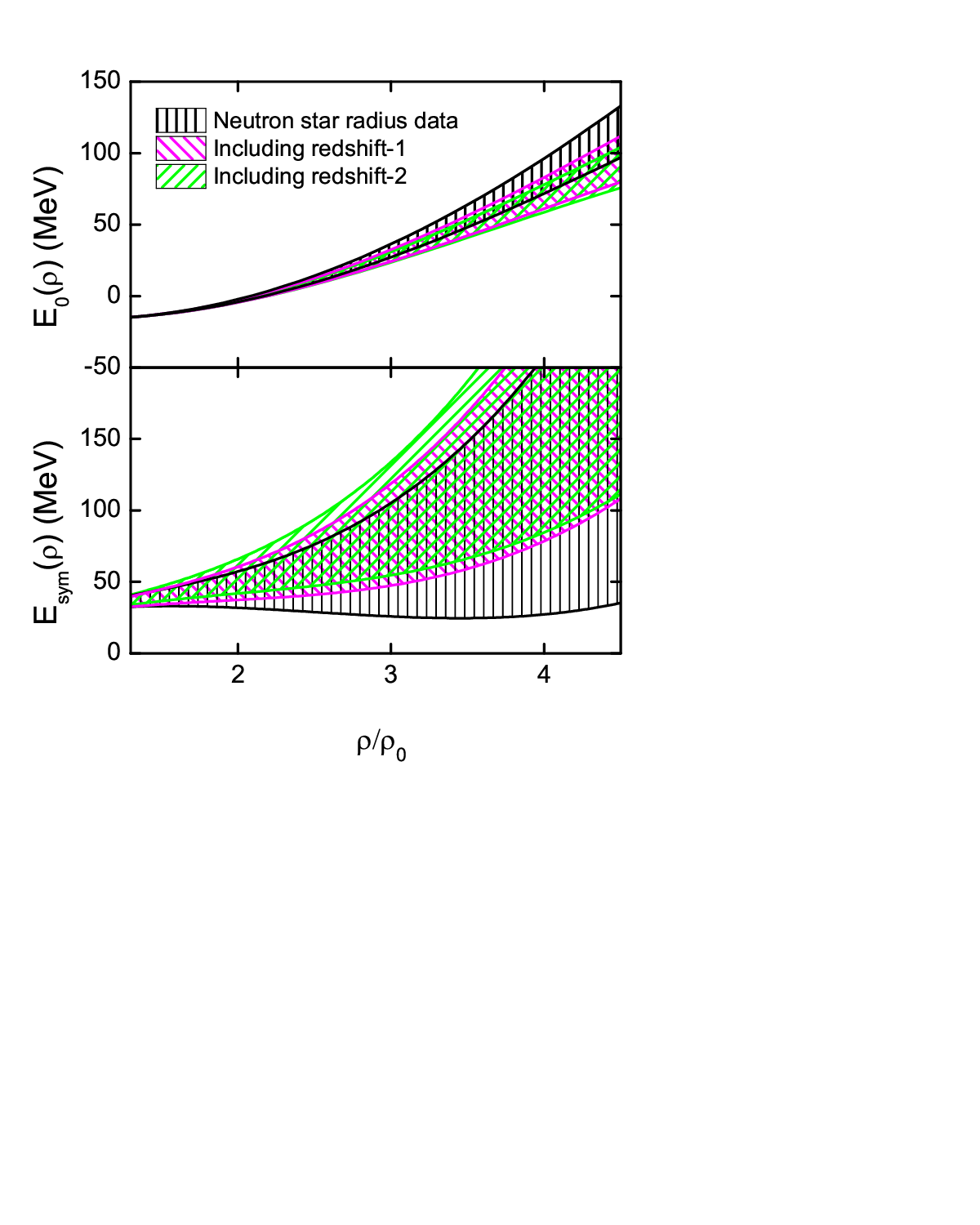}
  }
  \caption{(color online) The 68\% credible intervals of the binding energy in SNM (upper panel) and the nuclear symmetry energy (lower panel) as functions of the reduced density $\rho/\rho_0$ inferred from our Bayesian analyses with (magenta and green) and without (black) considering the gravitational redshift constraint from GS 1826-24. The results from the (non-)Gaussian form of the gravitational redshift distribution are represented by (green) magenta shadow marked as (Including redshift-2) Including redshift-1.} \label{e0esym}
\end{figure}

The upper and lower 68\% confidence boundaries of the binding energy $E_0(\rho)$ in SNM and the nuclear symmetry energy $E_{\mathrm{sym}}(\rho)$ corresponding to the results in Fig. \ref{6para} are displayed in Fig. \ref{e0esym}. As expected, the EOS of SNM exhibits a softening especially at high densities when the gravitational redshift data from GS 1826-24 is considered. On the other hand, the increased values of $K_{\text{sym}}$, $J_{\text{sym}}$, and $L$ lead to an enhancement of nuclear symmetry energy at supersaturation densities compared to the default calculations. Moreover, the 68\% confidence band of the symmetry energy at supersaturation densities becomes narrower thanks to the redshift constraint. Again, the effects especially on the high-density symmetry energy are stronger with the redshift-2 compared to the redshift-1.

To facilitate applications of the confidence boundaries of $E_0(\rho)$ and $E_{\mathrm{sym}}(\rho)$ in future studies, we have parameterized them as functions of the reduced density $x=\rho/\rho_0$. Considering the redshift data described with a Gaussian distribution (redshift-1) from GS 1826-24, their upper (u) and lower (l) 68\% confidence boundaries in units of MeV can be written respectively as
\begin{equation}\label{rzE068}
  E^u_0(\rho)=-1.11x^{3}+17.67x^2-32x-0.56,\\
  E^l_0(\rho)=-1.48x^{3}+17.44x^2-30.44x-1.52,
\end{equation}
and
\begin{equation}\label{rzEsym68}
  E^u_{\mathrm{sym}}(\rho)=4.94x^{3}-14.26x^2+35.7x+6.62,\\
  E^l_{\mathrm{sym}}(\rho)=3.24x^{3}-18.61x^2+41.5x+2.87.
\end{equation}
For the redshift data described with a non-Gaussian distribution (redshift-2) from GS 1826-24, they can be written respectively as
\begin{equation}\label{rzE068}
  E^u_0(\rho)=-1.3x^{3}+18.22x^2-32.56x-0.37,\\
  E^l_0(\rho)=-1.6x^{3}+17.93x^2-31.04x-1.28,
\end{equation}
and
\begin{equation}\label{rzEsym68}
  E^u_{\mathrm{sym}}(\rho)=4.94x^{3}-12.15x^2+34.81x+5.4,\\
  E^l_{\mathrm{sym}}(\rho)=2.87x^{3}-17.28x^2+44.61x-1.2.
\end{equation}

The results from using only the NS radius data can be written as
\begin{equation}\label{rE068}
  E^u_0(\rho)=-0.62x^{3}+16.19x^2-30.52x-1.05,\\
  E^l_0(\rho)=-1.05x^{3}+16.04x^2-28.93x-2.06,
\end{equation}
and
\begin{equation}\label{rEsym68}
  E^u_{\mathrm{sym}}(\rho)=4.94x^{3}-17.04x^2+39.26x+7.34,\\
  E^l_{\mathrm{sym}}(\rho)=2.47x^{3}-18.52x^2+39.63x+6.92.
\end{equation}

\begin{figure}[ht]
  \centering
 \resizebox{0.6\textwidth}{!}{
  \includegraphics[bb=10 200 520 550]{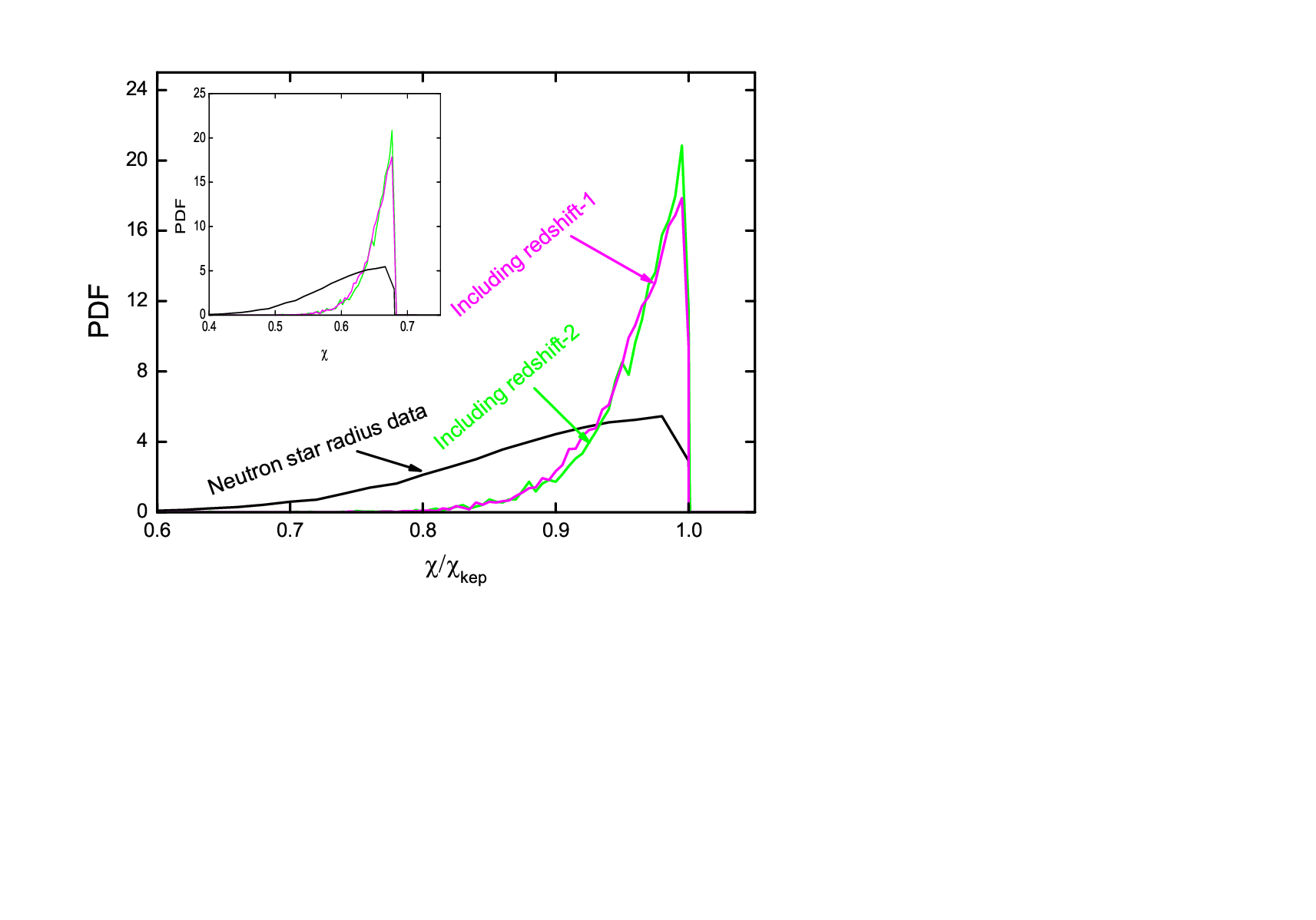}
  }
  \caption{(color online) Posterior probability distribution functions of the reduced dimensionless spin magnitude $\chi/\chi_{\rm kep}$ of a uniformly rotating star derived from the combined dataset of neutron star radius and gravitational redshift measurements (magenta and green) in comparison with the results (black) using solely the neutron star radius data listed in Table \ref{tab-data}. The inset shows the results of the dimensionless spin magnitude by adopting $\chi_{\mathrm{kep}}=0.68$ \cite{most2020Lower,koliogiannis2021Neutron}. The results from the combined dataset using the (non-)Gaussian distribution for the gravitational redshift measurements are shown by the (green) magenta curves marked as (Including redshift-2) Including redshift-1.}\label{x25}
\end{figure}

\begin{figure}[ht]
\vspace{-0.6cm}
\begin{center}
 \resizebox{0.49\textwidth}{!}{
  \includegraphics[width=15cm,height=15cm]{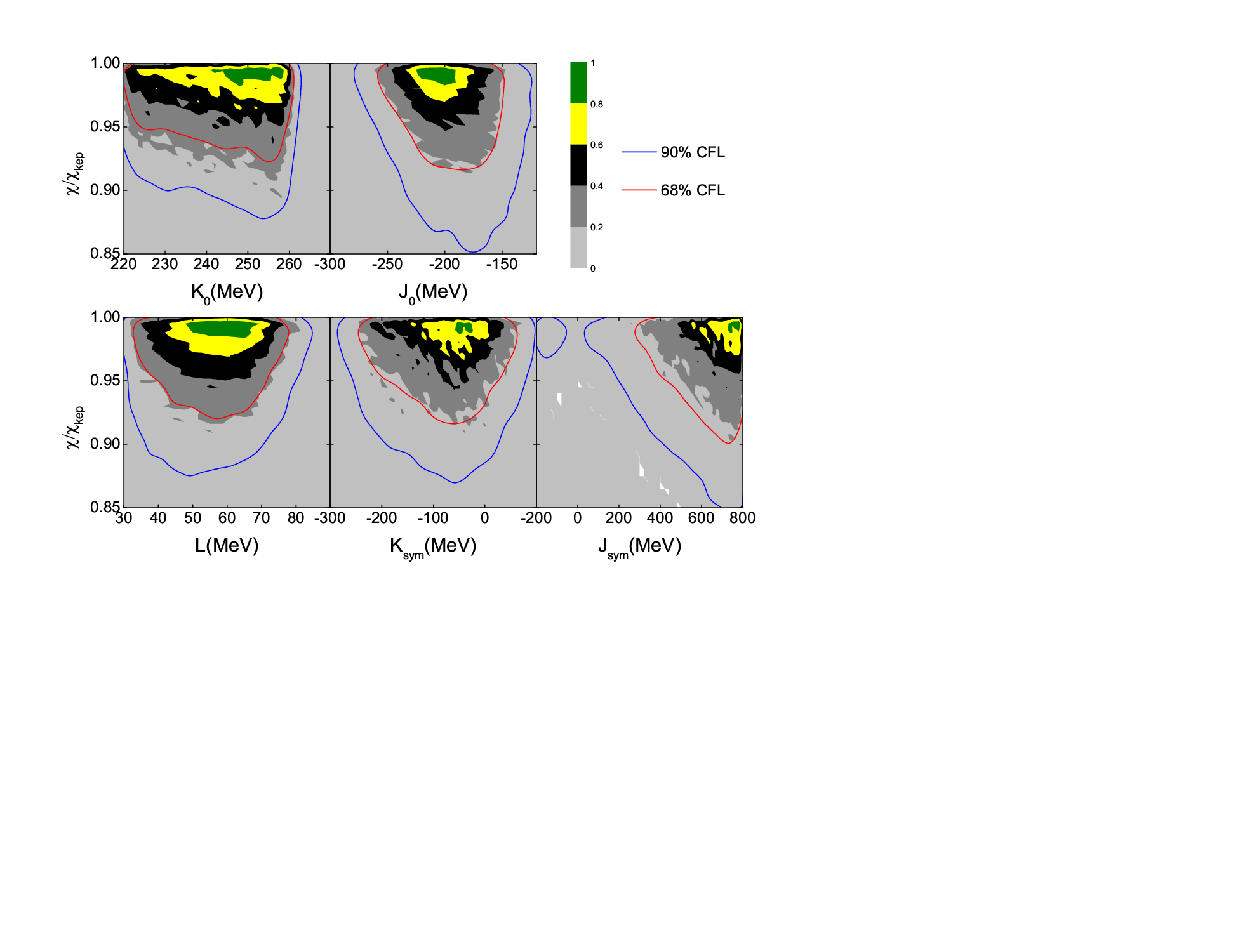}
  }
  \hspace{-2cm}
   \resizebox{0.49\textwidth}{!}{
  \includegraphics[width=15cm,height=15cm]{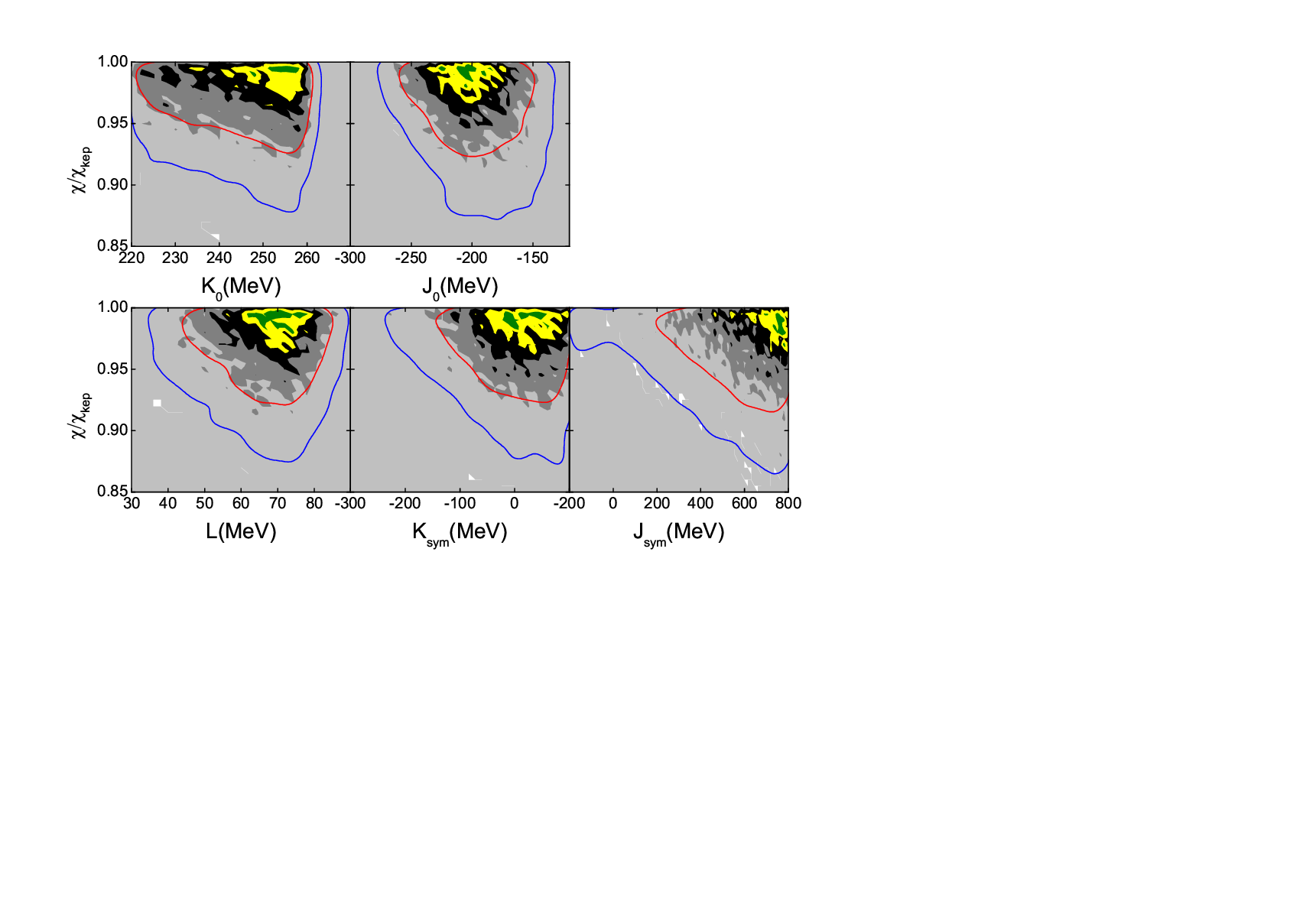}
  }

   \vspace{-3.5cm}
  \caption{(color online) Posterior correlation functions among the EOS parameters and the reduced dimensionless spin magnitude $\chi/\chi_{\mathrm{kep}}$ of a uniformly rotating NS to make its mass compatible with that of the second component of the GW190814 in our Bayesian analyses incorporating the gravitational redshift data with a Gaussian distribution (redshift-1) on the left and a non-Gaussian (redshift-2) on the right from GS1826-24, respectively. The blue and red curves represent the 90\% and 68\% confidence intervals, respectively.}\label{cor}
\end{center}
\end{figure}
\subsection{Impact on the spin of GW190814's minor component as a candidate of the most massive pulsar}
We now examine the impact of the gravitational redshift constraint from GS 1826-24 on the spin of GW190814's minor component as a candidate of the most massive pulsar. Shown in Fig. \ref{x25} are the PDFs of the reduced dimensionless spin magnitude $\chi/\chi_{\rm kep}$ with (magenta and green) and without (black) considering the redshift constraint from GS 1826-24. As mentioned earlier, these PDFs are inferred from our Bayesian analysis utilizing the likelihood function $P(D|{\cal M})= P(D|{\cal M})_{\rm base} \times P_\mathrm{z} \times P_\mathrm{M}$. The PDFs of the 6 EOS parameters remain almost identical to those presented in Fig.\ \ref{6para} and are thus not shown again here. It is seen that upon incorporating the gravitational redshift data from GS 1826-24, the necessary spin $\chi/\chi_{\rm kep}$ experiences a substantial increase. Moreover, its PDF is much narrower compared to the calculation solely relying on the NS radius data. A strong preference is indicated towards a larger spin parameter $\chi$ nearing its critical value $\chi_{\rm kep}$ where the pulsar becomes rotationally unstable, suggesting the second component $m_2$ of GW190814 is unlikely the most massive NS observed so far. It is interesting to note that while the change in PDF of $\chi/\chi_{\rm kep}$ is generally small, the redshift-2 prefers even more strongly the $\chi/\chi_{\rm kep}$ near its Keppler limit. This is simply because it prefers a smaller redshift, thus a smaller M$_{\rm TOV}$ as discussed earlier. Thus, to rotationally support the $m_2$ as a stable pulsar requires even faster rotations compared to the case with redshift-1. For a comparison, the inset in Fig. \ref{x25} displays the PDF of $\chi$ utilizing the value $\chi_{\text{kep}} = 0.68$ reported in Refs. \cite{most2020Lower,koliogiannis2021Neutron}. Our findings are consistent with the range $0.49_{-0.05}^{+0.08} \leq \chi \leq 0.68_{-0.05}^{+0.11}$ found in Refs. \cite{most2020Lower, koliogiannis2021Neutron} assuming M$^{\text{max}}_{\text{TOV}} < 2.3 M_{\odot}$ and $|\chi| \leq$ 0.81 found in Ref. \cite{annala2022multimessenger}, particularly when considering the gravitational redshift data. The latter naturally limits the maximum value of M$^{\text{max}}_{\text{TOV}}$ that is not set {\it apriori} in our analyses.

The posterior pairwise correlations between the EOS parameters and the spin parameter $\chi/\chi_{\mathrm{kep}}$ calculated by using the redshift-1 and redshift-2 are depicted in the left and right windows of Figure \ref{cor}, respectively. According to Eq. (\ref{breu and rezolla:universal relation}), the spin parameter $\chi/\chi_{\mathrm{kep}}$ and M$_{\mathrm{TOV}}$ are compensatory to produce an identical rotational mass $M_{\mathrm{max}}^{\mathrm{rot}}$ for pulsars. Thus, those EOS parameters (especially $J_0$ and $J_{\mathrm{sym}}$) leading to an increase of M$_{\mathrm{TOV}}$ as shown in the right panel of Fig. \ref{MmaxRmax} are expected to be negatively correlated with the spin parameter $\chi/\chi_{\mathrm{kep}}$. Indeed, this is the case, albeit weak.
For example, a negative correlation between $\chi/\chi_{\mathrm{kep}}$ and $J_0$ is readily apparent. Since $J_{\mathrm{sym}}$ is positively correlated with M$_{\mathrm{TOV}}$ as shown in the right panel of Fig. \ref{MmaxRmax}, a negative correlation also exists between $\chi/\chi_{\mathrm{kep}}$ and $J_{\mathrm{sym}}$. In other words, the decline in M$_{\mathrm{TOV}}$ requires a smaller total pressure in NSs, necessitating a softer NS EOS. All EOS parameters that increase the NS pressure are generally expected to be anti-correlated with the spin parameter. Namely, a smaller M$_{\mathrm{TOV}}$ supported by softer nuclear EOS needs to rotate faster such that its centrifugal force is high enough to help support the pulsar as heavy as the minor component of GW190814.

The pairwise correlations $L-\chi$ and $K_{\mathrm{sym}}-\chi$ are expected to be weak as both $L$ and $K_{\mathrm{sym}}$ affect the M$_{\mathrm{TOV}}$ very weakly as shown in the right panel of Fig. \ref{MmaxRmax}. Moreover, they may be overtaken by the correlations of $L$ and $K_{\mathrm{sym}}$ with $J_0$ and $J_{\mathrm{sym}}$ that affect most strongly the M$_{\mathrm{TOV}}$ (subsequently the $\chi/\chi_{\mathrm{kep}}$). For example, as shown in the Appendix, $K_{\mathrm{sym}}$ is positively correlated with $J_0$ but negatively with $J_{\mathrm{sym}}$ and their strengths depend on whether the redshift-1 or redshift-2 is used.
Thus, the marginalized pairwise correlations between $L$, $K_{\mathrm{sym}}$ and the spin $\chi/\chi_{\mathrm{kep}}$ may deviate from the indication of Fig. \ref{MmaxRmax} where only one parameter is varied at a time while all others are fixed at their most probable values. It is seen from Fig. \ref{cor} that $K_{\mathrm{sym}}$ becomes slightly anti-correlated with $\chi/\chi_{\mathrm{kep}}$ with the redshift-2, while there is no obvious change for the $L-\chi$ correlation compared to the calculation with the redshift-1. Therefore, through both direct and indirect correlations with M$_{\rm TOV}$, the observed weak $L-\chi$ and $K_{\mathrm{sym}}-\chi$ correlations as well as their variations with the redshift constraint are physically meaningful, although their interpretations are not trivial especially because $\chi$ is inversely proportional to M$_{\rm TOV}$ for a given rotational mass M$_{\rm max}^{\rm rot}$ to be supported, while the redshift constraint is on the M$_{\mathrm{TOV}}/{\rm R}_{\rm max}$ ratio. 

Since the relevant EOS parameters affect M$_{\mathrm{TOV}}$ and ${\rm R}_{\rm max}$ differently, considering all the constraints and data, the resulting correlations especially the weak ones among the EOS parameters and $\chi$ are thus not all straightforward. To appreciate the complexity involved and better understand the results shown in Fig. \ref{cor}, in the following we make a few comments regarding the expected direct $K_{\mathrm{sym}}-\chi$ correlation as well as their indirect correlations through $J_{\rm sym}$ and $J_0$.
\begin{enumerate}
\item
As discussed and demonstrated earlier, the redshift-2 favoring relatively smaller redshift values compared to the redshift-1 require both $L$ and $K_{\mathrm{sym}}$ to become larger while the $J_0$ gets smaller. The increased $L$ and $K_{\mathrm{sym}}$ not only make $R_{\rm max}$ larger but also reduce the  M$_{\mathrm{TOV}}$. They have the same net effect on reducing the compactness at M$_{\mathrm{TOV}}$ as a reduced $J_0$ (which makes the M$_{\mathrm{TOV}}$ significantly smaller and the $R_{\rm max}$ appreciably larger as shown in the right panel of Fig.  \ref{MmaxRmax}). In short,  increasing $L$ and/or $K_{\mathrm{sym}}$ {\it alone} will reduce the M$_{\mathrm{TOV}}$ appreciably, albeit weak. Consequently, higher $\chi/\chi_{\mathrm{kep}}$ values will be required to raise rotationally the M$_{\mathrm{TOV}}$ to be compatible with the observed mass $m_2$ of the minor component of GW190814. Therefore, if $L$ and $K_{\mathrm{sym}}$ are not correlated with any other EOS parameters, they are expected to be directly correlated positively with $\chi/\chi_{\mathrm{kep}}$ especially when the redshift-2 is used.
\item
Increasing the value of  $L$ and/or $K_{\mathrm{sym}}$ while fixing all other EOS parameters stiffens the symmetry energy at low-intermediate densities (around $\rho_0-2\rho_0$).
Consequently, the $R_{\rm max}$ increases but the M$_{\mathrm{TOV}}$ decreases appreciably as shown in the right panel of Fig.  \ref{MmaxRmax}. 
All together, they make the compactness or redshift decrease coherently. Their opposite effects on M$_{\mathrm{TOV}}$ and $R_{\rm max}$ are related to the $E_{\rm sym}(\rho)\delta^2$ term in the binding energy of neutron-rich matter and the $\beta$-equilibrium condition. If the latter requires a smaller (larger) $\delta$ with a higher (lower) $E_{\rm sym}(\rho)$ at low densities, the opposite is true at high densities when the symmetry energy at $\rho_0$ is fixed \cite{Lat91}. Also, to minimize the total energy, wherever/whenever the symmetry energy is higher the isospin asymmetry $\delta$ there is lower. The resulting density profile $\delta(\rho)$ at $\beta$-equilibrium exhibits the well-known isospin fractionation phenomenon ($E_{\rm sym}(\rho_1)\delta_1\approx E_{\rm sym}(\rho_2)\delta_2$ for two regions of density $\rho_1$ and $\rho_2$ in $\beta$-equilibrium \cite{Pawel,Zhang23-EPJA}), see, e.g., Ref. \cite{LCK} for a review. Since the total pressure in neutron stars depends on the SNM EOS $E_0(\rho)$, symmetry energy $E_{\mathrm{sym}}(\rho)$ as well as the isospin profile $\delta(\rho)$, see, e.g., Ref. \cite{Lat-sci} for a review, a stiffened $E_{\rm sym}(\rho)$ at low-intermediate densities necessary for increasing the $R_{\rm max}$ (note that the radius is determined by the radial coordinate where the pressure is zero while the M$_{\mathrm{TOV}}$ is determined by the maximum central pressure at high densities) may require simultaneously a softened high-density $E_{\rm sym}(\rho)$ (thus a reduced $J_{\rm sym}$ as shown in Fig. \ref{6para} with redshift-2 compared to that with redshift-1). 
In fact, mathematically one expects any two adjacent parameters (e.g., $K_0$ and $J_0$, L and $K_{\mathrm{sym}}$, $K_{\mathrm{sym}}$ and $J_{\mathrm{sym}}$) in parameterizing respectively $E_0(\rho)$ and $E_{\mathrm{sym}}(\rho)$ to be anti-correlated as discussed in more detail in the Appendix. 
The $E_{\rm sym}(\rho)$ above about $3\rho_0$ is controlled by the  $J_{\rm sym}$ parameter which monotonically increases both M$_{\mathrm{TOV}}$ and ${\rm R}_{\rm max}$ as shown in Fig. \ref{MmaxRmax}, leading directly to a negative $J_{\mathrm{sym}}-\chi$ correlation. Indirectly, then it would contribute to a positive $K_{\mathrm{sym}}-\chi$ correlation as $K_{\mathrm{sym}}$ and $J_{\mathrm{sym}}$ are always clearly anti-correlated.
\item As shown in the Appendix, to reproduce the radius data under the constraints discussed earlier, the $K_{\mathrm{sym}}$ is clearly positively correlated with $J_0$ in all cases, while the situation for L is less clear. Considering the competitions among the individual correlations discussed above, if the $K_{\mathrm{sym}}-J_0$ correlation dominates, $K_{\mathrm{sym}}$ is finally expected to be weakly anti-correlated with $\chi/\chi_{\mathrm{kep}}$ as observed in Fig.\ \ref{cor}.
\end{enumerate}

\section{Summary and Conclusions}\label{summary}
In summary, unprecedented high-precision nuclear mass measurements around the rp process waiting-point nucleus $^{64}$Ge have led recently to a significant revision of the surface gravitational redshift of the NS in GS 1826-24 by re-fitting its X-ray light curves using MESA. The NS compactness $\xi$ is constrained to be between 0.244 and 0.342 at 95\% confidence level and its upper boundary is significantly smaller than the maximum $\xi$ previously known. These findings call for a timely investigation of their impact on the EOS of supradense neutron-rich matter which is still poorly known but fundamentally important for resolving many critical issues in both astrophysics and nuclear physics. Within a comprehensive Bayesian statistical framework incorporating the newly revised redshift data from GS 1826-24, we studied the impact of the latter on the EOS of supradense neutron-rich matter and the necessary spin rate for GW190814's minor to be the most massive stable pulsar observed so far.

We found that the high-density SNM EOS has to be softened significantly compared to our prior knowledge from earlier analyses. On the contrary, the symmetry energy at supersaturation densities has to be stiffened significantly. In particular, the skewness $J_0$ characterizing the stiffness of high-density SNM decreases significantly compared to its fiducial value, while the slope $L$, curvature $K_{\rm{sym}}$, and skewness $J_{\rm{sym}}$ of nuclear symmetry energy all increase appreciably. Using the M$_{\rm TOV}$ constrained by the revised redshift data from GS 1826-24, the most probable spin rate for GW190814's minor $m_2$ to be a stable pulsar is found to be very close to its mass-shedding limit, indicating that the $m_2$ is unlikely the most massive pulsar observed so far.

\appendix
\section{Correlations among the EOS parameters from Bayesian analyses with and without including the gravitational redshift data}
\begin{figure}[ht]
\vspace{-0.6cm}
\begin{center}
 \resizebox{1.2\textwidth}{!}{
  \includegraphics[width=10cm,height=10cm]{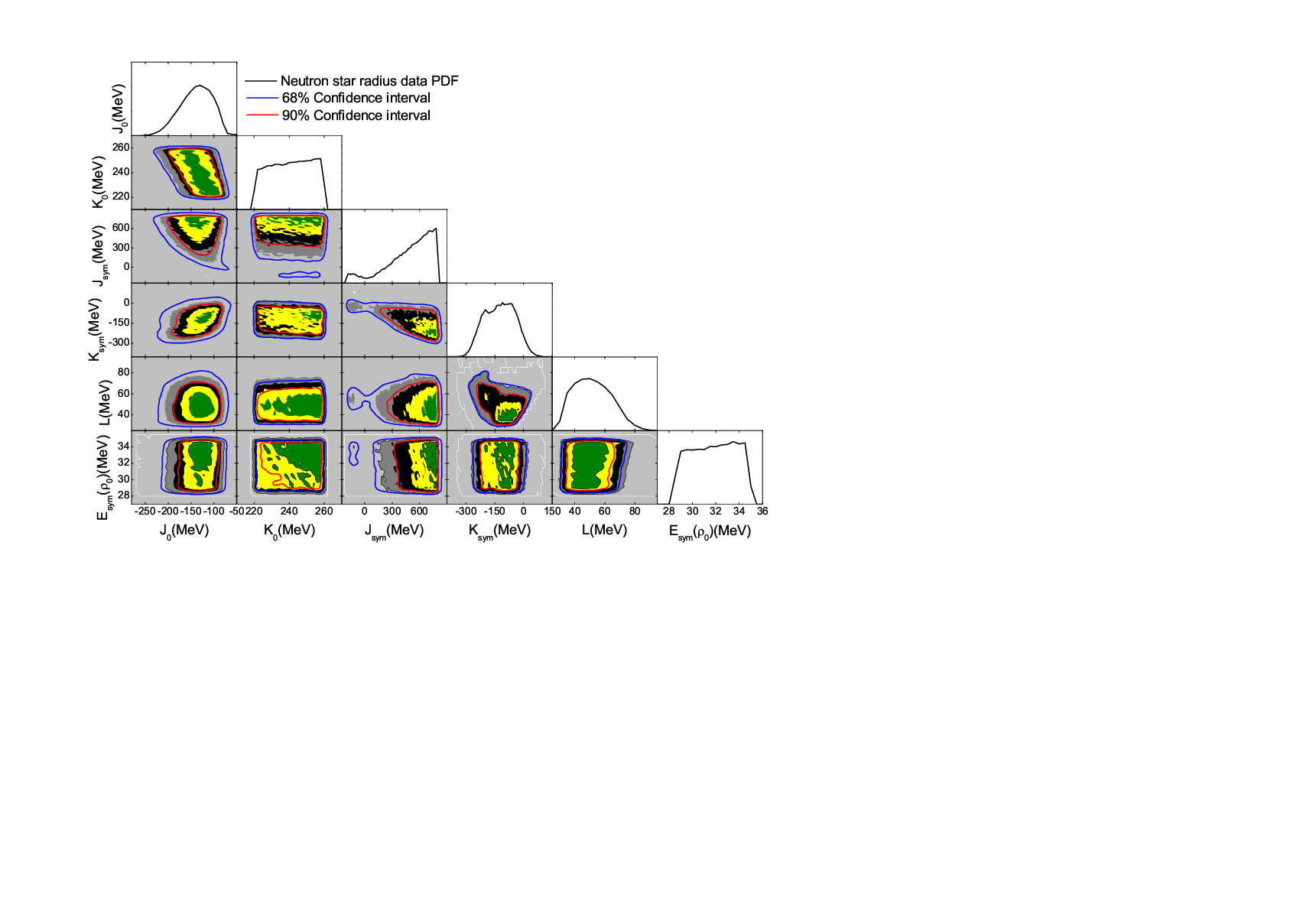}
  }
   \vspace{-9.cm}
  \caption{(color online) Posterior correlation functions among the EOS parameters in our Bayesian analyses based solely on the NS radius data. The blue and red curves represent the 90\% and 68\% confidence intervals, respectively.}\label{6paracor-ns}
\end{center}
\end{figure}
\begin{figure}[ht]
\vspace{-0.6cm}
\begin{center}
 \resizebox{1.2\textwidth}{!}{
  \includegraphics[width=10cm,height=10cm]{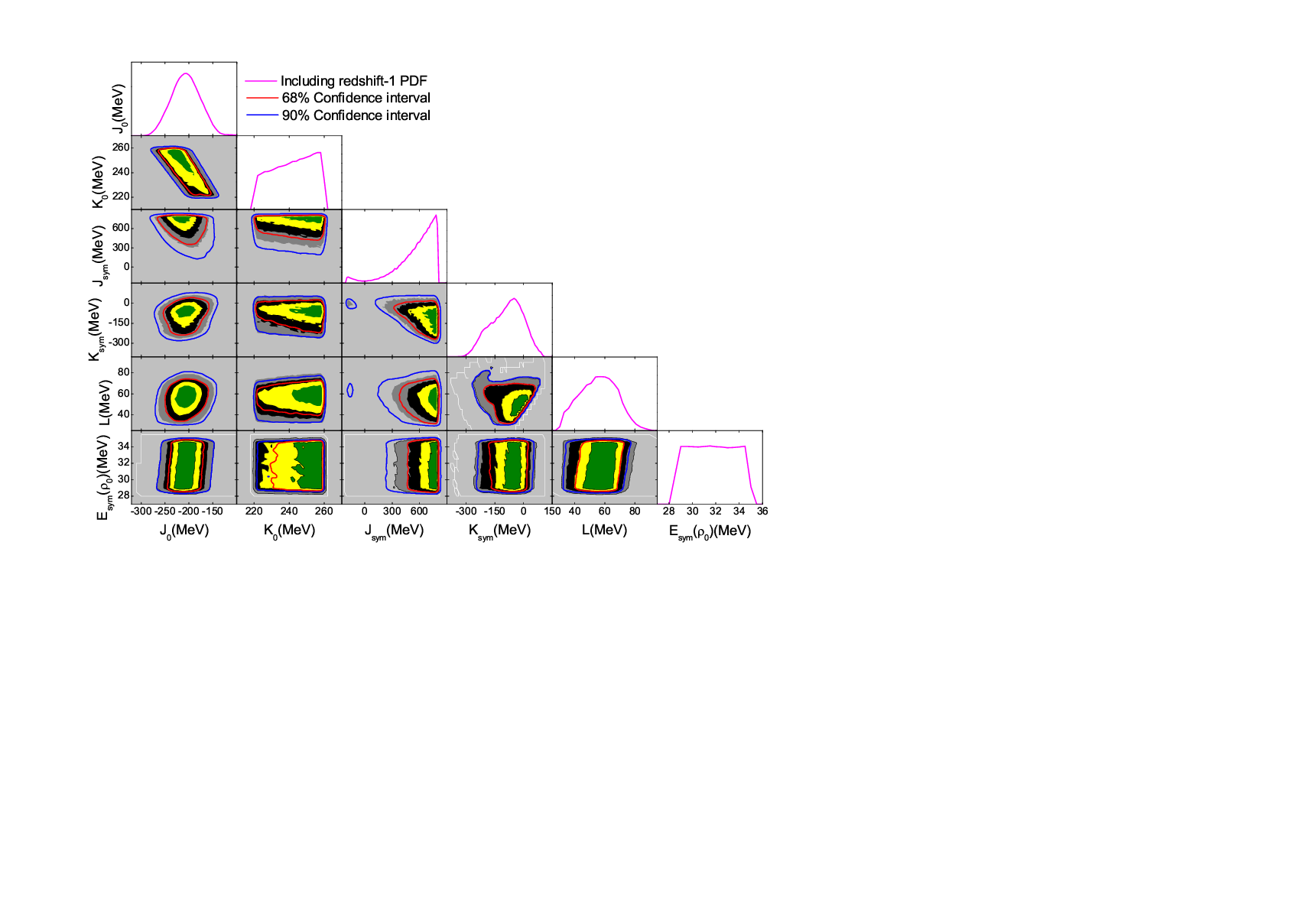}
  }
   \vspace{-9.cm}
  \caption{(color online) Same as Fig. \ref{6paracor-ns}, but showing the results from the Bayesian analyses including the gravitational redshift data with a Gaussian (redshift-1) distribution.}\label{6paracor-gau}
\end{center}
\end{figure}
\begin{figure}[ht]
\vspace{-0.6cm}
\begin{center}
 \resizebox{1.2\textwidth}{!}{
  \includegraphics[width=10cm,height=10cm]{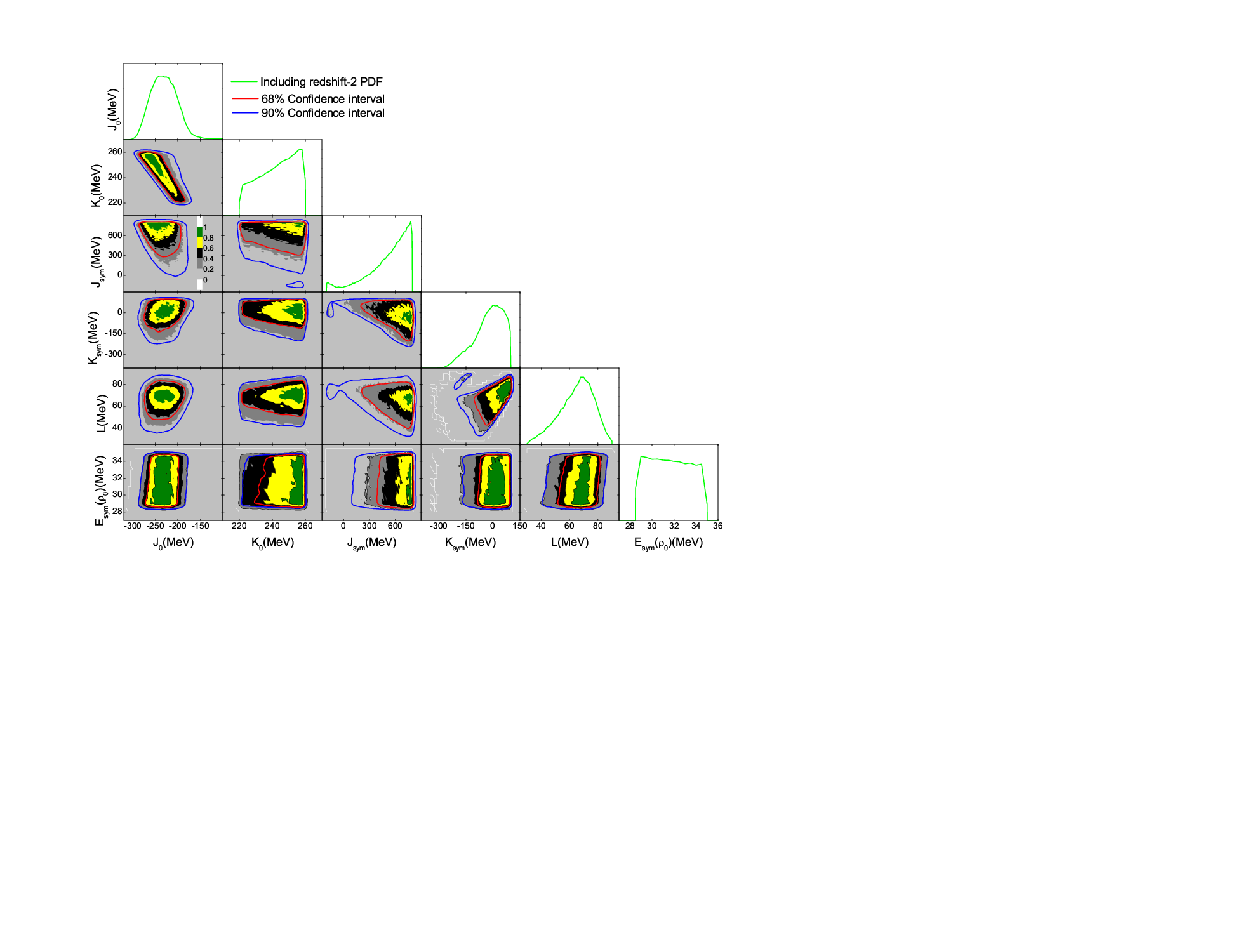}
  }
   \vspace{-9.cm}
  \caption{(color online) Same as Fig. \ref{6paracor-ns}, but showing the results from the Bayesian analyses including the gravitational redshift data with a non-Gaussian (redshift-2) distribution.}\label{6paracor-nogau}
\end{center}
\end{figure}

Corresponding to the PDFs of the six EOS parameters shown in Fig. \ref{6para} in the three cases investigated, shown in Figs. \ref{6paracor-ns}, \ref{6paracor-gau} and \ref{6paracor-nogau} are their pairwise correlations. We notice the following points.
\begin{enumerate}
\item Consistent with findings in our earlier Bayesian analyses \cite{xie2019bayesian,xie2020bayesian} and mathematical expectations, two adjacent parameters (e.g., $K_0$ and $J_0$, L and $K_{\mathrm{sym}}$, $K_{\mathrm{sym}}$ and $J_{\mathrm{sym}}$ in parameterizing respectively the SNM EOS $E_0(\rho)$ and symmetry energy $E_{\mathrm{sym}}(\rho)$ are anti-correlated, and the correlations between low-order and high-order (e.g., L and $J_{\mathrm{sym}}$) parameters are generally weak. The correlations between SNM EOS $E_0(\rho)$ and $E_{\mathrm{sym}}(\rho)$  parameters are introduced by both the radius data and the minimum M$_{\rm TOV}$ that all EOSs are required to produce. In particular,  $J_0$ and $J_{\mathrm{sym}}$ are anti-correlated as they both contribute positively to M$_{\rm TOV}$ as shown in the right window of Fig. \ref{MmaxRmax}. 
In our default analyses using the radius data only with several filters discussed earlier, it is seen that the $L$ and $K_{\mathrm{sym}}$ parameters are anti-correlated with each other, and $L$ has essentially no correlation with $J_0$ and $J_{\mathrm{sym}}$. On the other hand, $K_{\mathrm{sym}}$ is positively correlated with $J_0$ but negatively with $J_{\mathrm{sym}}$. Applying the redshift-1 z-distribution to the M$_{\mathrm{TOV}}$ configuration does not change qualitatively any of these features. This is mainly because the redshift-1 (compared to redshift-2) has a more extended high-z tail. It is thus relatively less restrictive as our default calculations have no limit on the redshift z until the causality line.
\item Particularly interestingly, applying the redshift-2 z-distribution that is more restrictive compared to the redshift-1 on the allowed high-z values (namely, the redshift-2 requires effectively smaller compactness at the M$_{\mathrm{TOV}}$ configuration), the $L$ and $K_{\mathrm{sym}}$ become positively correlated with each other. Moreover, the positive correlation between $J_0$ and $K_{\mathrm{sym}}$ is weakened. These features are all understandable. Applying the redshift-2 z-distribution to the M$_{\mathrm{TOV}}$ configuration requires effectively a reduction of $\xi_{\rm max}=$M$_{\mathrm{max}}/{\rm R}_{\rm max}$ as we discussed earlier.  As shown in Fig.\ \ref{MmaxRmax}, increasing $J_0$ raises M$_{\mathrm{TOV}}$ but lowers R$_{\mathrm{max}}$ (they increases coherently the $\xi_{\rm max}$). On the other hand, increasing either $L$ and/or $K_{\mathrm{sym}}$ lowers M$_{\mathrm{TOV}}$ but raises the R$_{\mathrm{max}}$ appreciably (they decreases coherently the $\xi_{\rm max}$) by keeping all other parameters fixed at their most probable values. Therefore, the smaller compactness $\xi_{\rm max}$ with the redshift-2 data requires $L$ and/or $K_{\mathrm{sym}}$ to increase, while the $J_0$ decrease. Consequently,  the $L$ and $K_{\mathrm{sym}}$ become positively correlated in the case of using the redshift-2 z-distribution. Simultaneously, their correlations with $J_0$ become weaker compared to the calculations without considering the redshift data.
\end{enumerate}

\noindent{\bf Acknowledgement:} We would like to thank Bao-Jun Cai, James Lattimer and Hendrik Schatz for helpful discussions.
WJX is supported in part by the Shanxi Provincial Foundation for Returned Overseas Scholars under Grant No 20220037, the Natural Science Foundation of Shanxi Province under Grant No 20210302123085, the Open Project of Guangxi Key Laboratory of Nuclear Physics and Nuclear Technology, No. NLK2023-03 and the Central Government Guidance Funds for Local Scientific and Technological Development, China (No. Guike ZY22096024). BAL is supported in part by the U.S. Department of Energy, Office of Science, under Award Number DE-SC0013702, the CUSTIPEN (China-U.S. Theory Institute for Physics with Exotic Nuclei) under the US Department of Energy Grant No. DE-SC0009971. NBZ is supported in part by the National Natural Science Foundation of China under Grant No. 12375120, the Fundamental Research Funds for the Central Universities under Grant No. 2242024RCB0013 and the Start-up Research Fund of Southeast University under Grant No. RF1028623060.


\end{document}